\documentclass{aa}
\usepackage{txfonts}
\usepackage[version=4]{mhchem}
\usepackage[colorlinks, linkcolor=blue, citecolor=blue]{hyperref}
\usepackage{subfigure}
\usepackage{graphicx}   
\graphicspath{{./figs/}}

\newcounter{chem}
\newcounter{temp}

\newenvironment{chequation}{%
\setcounter{temp}{\value{equation}}%
\setcounter{equation}{\value{chem}}%
\renewcommand{\theequation}{R\arabic{equation}}%
}{%
\setcounter{chem}{\value{equation}}%
\setcounter{equation}{\value{temp}}%
}

\begin{document}

\title{Astrochemical effect of the fundamental grain surface processes I. The diffusion of grain surface species and the pre-exponential factor}

\author{Long-Fei Chen\inst{1}
\and Donghui Quan\inst{1}
\and Jiao He\inst{2,3}
\and Yao Wang\inst{4}
\and Di Li\inst{5,1,6}
\and Thomas Henning\inst{3}
}

\institute{Research Center for Astronomical Computing, Zhejiang Laboratory, Hangzhou 311100, PR China; \email{donghui.quan@zhejianglab.com}
\and
College of Engineering Physics, Shenzhen Technology University, Shenzhen 518118, PR China; \email{hejiao@sztu.edu.cn}
\and
Max Planck Institute for Astronomy, K{\"o}nigstuhl 17, Heidelberg 69117, Germany;
\and
Purple Mountain Observatory, Chinese Academy of Sciences, 10 Yuanhua Road, Nanjing 210023, PR China
\and
National Astronomical Observatories, Chinese Academy of Sciences, 20A Datun Road, Chaoyang District, Beijing 100101, PR China
\and
Department of Astronomy, College of Physics and Electronic Engineering, Qilu Normal University, 2 Wenbo Road, Zhangqiu District, Jinan 250200, PR China
}

\authorrunning{Chen et al.}
\titlerunning{Astrochemical effect of the pre-exponential factor}

\date{Received date month year / Accepted date month year}

\abstract
{Thermal diffusion is one of the basic processes for the mobility and formation of species on cosmic dust grains. The rate of thermal diffusion is determined by the grain surface temperature, a pre-exponential factor, and an activation energy barrier for diffusion. Due to the lack of laboratory measurements on diffusion, prior astrochemical models usually assume that the diffusion pre-exponential factor is the same as that for desorption. This oversimplification may lead to an uncertainty in the model predictions. Recent laboratory measurements have found that the diffusion pre-exponential factor can differ from that for desorption by several orders of magnitude. However, the newly determined pre-exponential factor has not been tested in astrochemical models so far.}
{We aim to evaluate the effect of the newly experimentally measured diffusion pre-exponential factor on the chemistry under cold molecular cloud conditions.}
{We ran a set of parameters with different grain temperatures and diffusion barrier energies using the NAUTILUS astrochemical code and compared the molecular abundance between the models with the abundance obtained using the experimentally determined pre-exponential factor for diffusion and with the abundance obtained using the values commonly adopted in prior models.}
{We found that statistically, more than half of the total gas-phase and grain surface species are not affected by the new pre-exponential factor after a chemical evolution of 10$^5$ yr. The most abundant gas-phase \ce{CO} and grain surface water ice are not affected by the new pre-exponential factor. For the grain surface species that are affected, compared to the commonly adopted value of the pre-exponential factor for diffusion used in the chemical models, they could be either overproduced or underproduced with the lower diffusion pre-factor used in this work. The former case applies to radicals and the species that serve as reactants, while the latter case applies to complex organic molecules (COMs) on the grain and the species that rarely react with other species. Gas-phase species could also be affected due to the desorption of the grain surface species. The abundance of some gas-phase COMs could be varied by over one order of magnitude depending on the adopted grain surface temperature and/or the ratio of diffusion to desorption energy in the model. Key species whose diffusion pre-exponential factor significantly affects the model predictions were also evaluated, and these specie include \ce{CH3OH}, \ce{H2CO}, and \ce{NO}.}
{The results presented in this study show that the pre-exponential factor is one of the basic and important parameters in astrochemical models. It strongly affects the chemistry and should be determined carefully. More experiments to determine the diffusion of grain surface species are helpful for constraining their properties.}


\maketitle

\section{Introduction} \label{sec.intro}
Astrochemical models are useful tools for studying the observed molecules detected in various sources in the interstellar medium (ISM), such as molecular clouds and star formation regions. The results from astrochemical models can provide perspectives and predictions on the evolution of the molecules in the ISM. For example, the molecular abundances from astrochemical models can be compared with the observational abundances to reveal the origin and formation pathways of the observed molecules, such as complex organic molecules (COMs; \citet{Herbst2009}), and the chemistry in star formation regions \citep{Rivilla2017, Quenard2018, Zeng2021}. The predictions of the modeling results can also be used to guide the observations, such as in the quest for the prebiotic molecules \citep{Garrod2013, Jimenez-Serra2014, Suzuki2018, Zhao2021}.

To leverage the power of chemical models, their reliability should be examined extensively. The reliability and accuracy of the predictions of the chemical models depend on the initial assumptions and parameters used in the models. Due to the complex interactions among the chemical reactions and the different mechanisms considered in the models, many factors can affect the evolution of the chemistry. Thousands of reactions are included in the chemical reaction network, while many reactions may not have been experimentally studied \citep{Viti2020}. By considering the reaction rate uncertainties in the chemical models, \citet{Wakelam2005, Wakelam2006} found that critical species and reactions should be included in the chemical network to better produce model results when compared with observations. On the other hand, the loose constraints of desorption energies and diffusion energies, as well as their corresponding pre-exponential factors for grain surface species, are another source of uncertainty \citep{Penteado2017, Iqbal2018, Furuya2022}.

One of the basic and crucial mechanisms in the processes of the grain surface chemistry \citep{Hasegawa1992, Semenov2010} for the mobility and reactions of the surface species is thermal diffusion, especially for radicals, which diffuse and recombine to form new molecules. Grain surface species and their associated reactions strongly affect chemical evolution \citep{Chang2007, Cuppen2009, Ruaud2016}. Ice mantles that cover the grain surface are initially grown via the accretion of gas-phase atoms and molecules onto cosmic dusts \citep{Charnley2001}. The complexity of the ice species is further increased by the hydrogenation and recombination of radicals, as well as by the dissociation of larger species by UV photons or cosmic rays \citep{Theule2013, Enrique-Romero2022}. COMs could be formed in the ice mantles by the recombination of radicals \citep{Fedoseev2015, Chuang2017, He2022}. The ice mantle species are then released into the gas phase by thermal or various nonthermal desorption mechanisms. The large number of gas-phase COMs detected in the star formation regions of hot cores or corinos is generally thought to originate from the cold core conditions, where the COMs are buried in the ice mantles \citep{Oberg2010, Suzuki2018, vanGelder2020, Nazari2022}. Therefore, the evaluation of the basic grain surface processes is crucial for the grain surface chemistry and the explanation of the observations.

The diffusion and recombination of grain surface species are two basic but important chemical processes. The rate of the thermal diffusion depends on the diffusion energy barrier ($\mathrm{E_{dif}}$) of the species and the dust temperature ($\mathrm{T_d}$) and can be expressed according to the Arrhenius law by the following equation,

\begin{equation}
\mathrm{k = \nu_{\rm dif} ~ exp(-E_{\rm dif}/T_d)},
\label{eq.diffusion}
\end{equation}

where $\nu_{\rm dif}$ is the pre-exponential factor or vibrational frequency \citep{Hasegawa1992, Semenov2010}. Because it is generally challenging to measure the pre-exponential factor for diffusion in the laboratory, in astrochemical models it is a common practice to assume the same pre-exponential value as desorption. The value for desorption is usually estimated using the following expression: 

\begin{equation}
\mathrm{\nu_{des} = \sqrt{2 N_s k_B E_{\rm des} / \pi^2 m m_p}}, 
\label{eq.Hasegawa}
\end{equation}

where $N_s$ is the grain surface site density ($\approx$1.5 $\times$ 10$^{15}$ cm$^{-2}$), $k_B$ is the Boltzmann constant, $E_{\rm des}$ is the desorption energy of the species in Kelvin, $m$ is the molecular weight of the species in atomic mass units (amu), and $m_p$ is the mass of proton. The calculated values of the pre-exponential factor for desorption are generally in the range of 10$^{12}$--10$^{13}$ s$^{-1}$ for most species existing in the ISM. This value is also most commonly used in astrochemical models for both desorption and diffusion. 

In principle, diffusion and desorption are two different processes, and their pre-exponential factors are usually different. Experimental evidence has been reported to show that the pre-exponential factor for diffusion can be orders of magnitude different from that for desorption \citep[see e.g. ][and references therein]{Wang2002}. It is important to perform separate laboratory measurements on diffusion rather than assuming the same value as that for desorption. Due to technical difficulties in the measurements, much less is known about the diffusion pre-exponential factor than about desorption. For a review of the laboratory measurements on diffusion, we refer to the recent work by \citet{He2023}.  

In an effort to determine the pre-exponential factor and energy barrier values simultaneously for several astrochemically relevant molecules on the surface of amorphous solid water (ASW), \citet{He2018} designed a set of laboratory experiments on diffusion that used the shifting of dangling OH bonds in ASW to trace diffusion. They found that the values of the diffusion pre-exponential factor could range from 10$^8$ to 10$^9$ s$^{-1}$ for \ce{CO}, \ce{N2}, \ce{O2}, and \ce{CH4}. These values are 3 to 4 orders of magnitude lower than the adopted values for desorption in the modeling works mentioned above. Since \ce{CO}, \ce{N2}, \ce{O2}, and \ce{CH4} have similar volatilities, the question remained whether this pre-exponential factor value is generally applicable to other molecules as well. With this in mind, more recently, \citet{He2023} extended the laboratory measurements to a less volatile molecule, \ce{CO2}, and found a similar pre-exponential factor value for diffusion of \ce{CO2} on nonporous ASW. It is therefore plausible that the lower pre-exponential factor value for diffusion could be applicable to a wider range of molecules on ASW. In this study, we test this new diffusion pre-exponential factor (referred to as pre-factor hereafter if not especially stated) value on the effect of astrochemistry.

Equation (\ref{eq.diffusion}) shows that the diffusion rate is proportional to the value of the pre-factor used. It can therefore be expected that the different chosen values of the pre-factor could have notable impact on the evolution of chemistry. \citet{Acharyya2022} tested the effect of the pre-factor on the chemistry for the grain surface \ce{CO} alone, and the results showed no difference among models with different values of the pre-factor for \ce{CO}. In this paper, we systematically investigate the effect of the pre-factor on the chemistry of all species in the chemical network, with a focus on the chemical discrepancy of the astrochemical models. The effect of the diffuse energy of a species and grain temperature on the astrochemical models is also discussed. Section \ref{sec.model} presents the description of the astrochemical models. Section \ref{sec.results} shows a full comparison of the modeled results of different groups of gas-phase species as well as grain surface species. Section \ref{sec.discussions} discusses the effects of the pre-factor, and Section \ref{sec.summary} summarizes our conclusions.

\section{Model description} \label{sec.model}
Diverse values of the pre-factor can be derived by different theoretical methods and experiments. The theoretical calculations of the pre-factor were presented in \citet{Hasegawa1992}, and we refer to the discussions in Section \ref{sec.discussions}. In this study, we distinguished the pre-factor for diffusion and the pre-factor for desorption in the models, and focused on the chemical discrepancy of the astrochemical models when a lower value of the pre-factor for diffusion was used, as indicated by the experimental results by \citet{He2018,He2023}, while the pre-factor for desorption was treated as usual. We also distinguished the differences between the diffusion pre-factor and desorption pre-factor in the reaction-diffusion competition mechanism, where there is a competition among reaction, hopping, and evaporation processes \citep{Ruaud2016}. The diffusion pre-factor is used to calculate the probability of reaction and hopping rate, and the desorption pre-factor is used to calculate the evaporation rate.

We used the public version of the gas-grain astrochemical simulation code NAUTILUS \footnote{https://kida.astrochem-tools.org/codes.html} \citep{Ruaud2016} to investigate the effect of the pre-factor on the modeling results. A three-phase modeling, which includes the gas phase, grain surface, and ice mantles, was considered. We ran simulations under the typical cold molecular cloud conditions with a visual extinction of 15 mag and a hydrogen number density of 2 $\times$ 10$^4$ cm$^{-3}$. The initial elemental abundances used in the models were the same as in \citet{Semenov2010}. All other physical and chemical parameters were set to default values, except for the changes described as follows.

The thermal diffusion rate of Eq.~(\ref{eq.diffusion}) shows that the rate depends on the pre-factor, the grain temperature (T$_d$), and the diffusion barrier $E_{\rm dif}$. We therefore ran a series of parameters in which T$_d$ changed from 10 to 20 K (but the gas temperature was set to be the same as the grain temperature). 
For the species for which both diffusion energy barrier and pre-factor have been measured in the laboratory, we adopted the measured values in the model. The $E_{\rm dif}$ value of most species is poorly constrained, and it is usually assumed that the ratio $E_{\rm dif}$/$E_{\rm des}$ is constant. Experiments on the measurements of the diffusion of volatiles in ASW under ISM conditions showed that this value can range from 0.3 to 0.5 \citep{He2018}, depending on whether $E_{\rm des}$ is considered to be the value corresponding to the extremely low coverage or monolayer coverage. For three-phase astrochemical models, a higher value of 0.8 was used for the species below the topmost two layers of the ice mantles, considering that the species have a higher inertia in the deeper ice mantles than in the surface ice mantles. We used three values, 0.3, 0.4, and 0.5, of the $E_{\rm dif}$/$E_{\rm des}$ ratio for the grain surface species to test the impact of the diffusion barrier on the chemistry.
Although the constant ratio assumption may not be valid, this could be considered as a necessary compromise before more extensive laboratory data on diffusion are available. 

Equation (\ref{eq.diffusion}) shows that the diffusion rate is negatively related to the diffusion barrier. We therefore expect that a high value of ${E_{\rm dif}}$ might compensate for the impact of the pre-factor on the chemistry. For $E_{\rm des}$ in the models, we used the recently updated set of the desorption energy described in \citet{Wakelam2017}. For \ce{CO}, \ce{N2}, \ce{O2}, \ce{CH4}, \ce{H2CO}, \ce{CH3OH}, and \ce{CO2}, their desorption energy and diffusion barrier were taken from the experimental results \citep{He2018, He2022, He2023}. For the radicals of \ce{HCO}, \ce{CH2OH}, and \ce{CH3O}, their desorption energies were calculated as shown in Table \ref{tab.ED}.

Radicals are essential for the formation of COMs in the grain ice mantles, and the effect of diffusion pre-factor has a different impact on the radicals and COMs. Therefore, based on NAUTILUS, we extended the chemical reaction network to include \ce{CH3OCH2OH}, \ce{(CH3O)2}, \ce{(CH2OH)2}, and other related species and reactions from \citet{Garrod2013}. These COMs could be produced by the recombination of radicals such as \ce{CH3O}, \ce{CH2OH}, and \ce{NH2}. In total, there were 655 gas-phase species, 281 grain surface species, and 12549 gas-grain reactions in the network we used.

Hereafter, the reference models M1 adopt the common value of the pre-factor, and the new models M2 use updated pre-factor values of 10$^9$ s$^{-1}$ with a uniform assumption. For each of the models, different grain temperature and $E_{\rm dif}$/$E_{\rm des}$ ratios were chosen to test their effect on the pre-factor.

\begin{table*}[htbp]
\centering
\caption{Desorption energy ($E_{\rm des}$) and diffusion barrier ($E_{\rm dif}$) of the species obtained from experiments.}
\label{tab.ED}
\begin{tabular}{llll}
\hline
\hline
Species    & $E_{\rm des}$ (K)  & $E_{\rm dif}$ (K) & Reference \\
\hline
\ce{H}     & 650                & 230               & \citet{Cuppen2007} \\
\ce{H2}    & 440                & 220               & \citet{Cuppen2007} \\
\ce{CO}    & 1600               & 490               & \citet{He2018} \\
\ce{N2}    & 1320               & 447               & \citet{He2018} \\
\ce{O2}    & 1520               & 446               & \citet{He2018} \\
\ce{CH4}   & 1600               & 547               & \citet{He2018} \\
\ce{H2CO}  & 2980               & -$^a$             & \citet{He2022} \\
\ce{CH3OH} & 4400               & -                 & \citet{He2022} \\
\ce{CO2}   & 2600               & 1300              & \citet{Wakelam2017, He2023} \\
\ce{HCO}   & 2290$^b$           & -                 & \\
\ce{CH2OH} & 3750$^c$           & -                 & \\
\ce{CH3O}  & 3630$^d$           & -                 & \\
\hline
\end{tabular}\\
Notes\\
$^a$Determined by the $E_{\rm dif}$/$E_{\rm des}$ ratio.\\
$^b$Calculated from the desorption energy of \ce{CO} and \ce{H2CO}: ($E_{\rm des,CO}$ + $E_{\rm des,H2CO}$)/2.\\
$^c$Calculated from the desorption energy of \ce{CH3OH} and \ce{H}: ($E_{\rm des,CH3OH}$ - $E_{\rm des,H}$)/2.\\
$^d$Calculated from the desorption energy of \ce{H2CO} and \ce{H}: ($E_{\rm des,H2CO}$ + $E_{\rm des,H}$)/2.
\end{table*}

\section{Results and analysis} \label{sec.results}
To comprehensively illustrate the difference in the results between the reference model M1 and the new model M2, we divided the species in the network into six groups, namely the major ice components (e.g.,~\ce{H2O} and \ce{CO}), the minor ice components (e.g.,~\ce{N2} and \ce{O2}), the grain radicals (e.g.,~\ce{OH} and \ce{CH2OH}), the grain COMs (e.g.,~\ce{CH3OH} and \ce{C2H5OH}), the simple gas-phase species (e.g.,~\ce{CO} and \ce{CO2}), and the gas-phase COMs, for a thorough analysis with representative species and to evaluate the impact of the diffusion pre-factor on their chemistry.

\subsection{Overview of the two models}\label{sec.overview}
Thermal diffusion is a grain surface process. It is therefore expected that the influence of the changes of the pre-factor affect the grain surface species and their chemistry first. Consequently, due to the interactions of gas-phase and grain species by adsorption, thermal desorption, and nonthermal desorption processes, all the species in the network could be affected. Thus, the evolution of all the species in the network, that is, their abundance as a function of time, could have a complex response due to the changes of the pre-factor.

\begin{figure*}[htbp]
\centering
\subfigure[Model: T$_d$ = 10 K, $E_{\rm dif}$/$E_{\rm des}$ = 0.5]{
    \includegraphics[scale=0.4]{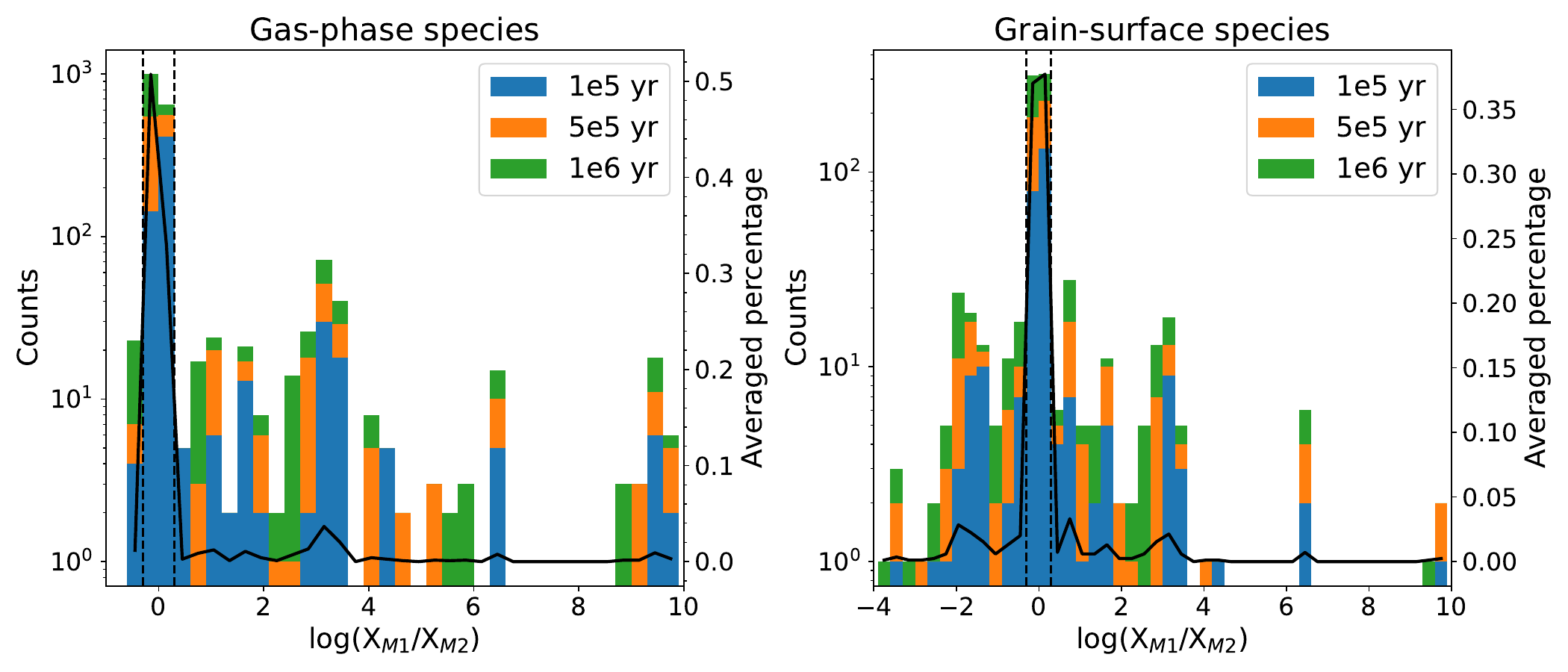}
    \label{fig.overview_hist.10,0.5}
}
\quad  
\subfigure[Model: T$_d$ = 20 K, $E_{\rm dif}$/$E_{\rm des}$ = 0.3]{
    \includegraphics[scale=0.4]{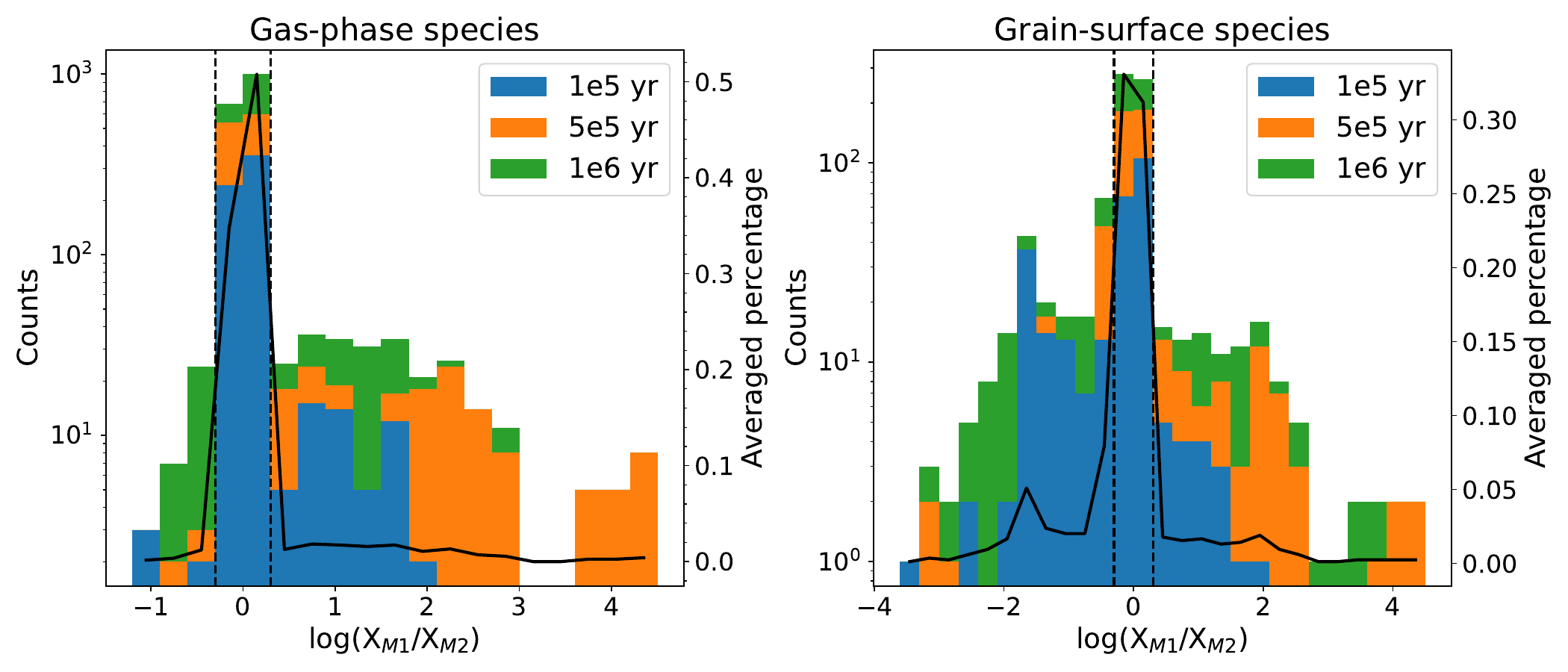}
    \label{fig.overview_hist.20,0.3}
}
\caption{Statistics of the logarithmic molecular abundance ratio of M1 and M2 for the gas-phase species and grain species at the three different evolutionary times. Panel (a) shows modeling parameters with a grain temperature T$_d$ = 10 K and $E_{\rm dif}$/$E_{\rm des}$ = 0.5, and panel (b) shows T$_d$ = 20 K and $E_{\rm dif}$/$E_{\rm des}$ = 0.3. X$_{M1}$ and X$_{M2}$ represent the abundance of a species in models M1 and M2, respectively. The solid lines represent the relative percentage of each bin with respect to the total number of the gas-phase or grain species averaged over the three different times. The width of each bin is 0.3.}
\label{fig.overview_hist}
\end{figure*}

We used the logarithmic ratio of the species abundance between models M1 and M2 to indicate the discrepancy of the two models. A positive value of the logarithmic abundance ratio represents an underproduction of species in M2, but an overproduction for species in M2 when the logarithmic abundance ratio is negative. We assumed that a value beyond the range of $\pm$0.3, which corresponds to a double difference between two models, is significant. Figure \ref{fig.overview_hist} shows the histogram comparison of this logarithmic molecular abundance ratio of models M1 and M2 for the gas-phase species and grain species at three different evolutionary times, which are 1 $\times$ 10$^5$ yr, 5 $\times$ 10$^5$ yr, and 1 $\times$ 10$^6$ yr. These times are selected based on the general evolutionary timescales of the molecular clouds. Panel (a) represents the results for the model with a grain temperature T$_d$ = 10 K and $E_{\rm dif}$/$E_{\rm des}$ = 0.5, and panel (b) represents the results for T$_d$ = 20 K and $E_{\rm dif}$/$E_{\rm des}$ = 0.3. As indicated in Eq.~(\ref{eq.diffusion}), the thermal diffusion rate is positively related to the grain temperature and negatively related to the diffusion barrier energy. Therefore, the above two sets of model parameters represent two extreme conditions that have the lowest and highest rate constants, respectively. In the figures, the counts are the absolute numbers of the species, and the solid line represents the relative percentage of each bin with respect to the total number of the gas-phase or grain species averaged over the three different times. The width of each bin in the histogram is 0.3. Therefore, the bins between the vertical dashed lines in the figures represent a difference smaller than 2 for the abundance of a species between M1 and M2.

Figure \ref{fig.overview_hist.10,0.5} shows that there are 84\% and 75\% species with respect to the total number of the gas-phase species and the grain species, respectively, that are within the double difference between M1 and M2. However, model M2 can underproduce the abundance of the gas-phase species by as low as 9 orders of magnitude compared with model M1 for several species. For the grain species, the logarithmic difference between M1 and M2 is distributed from -4 to 10, which means that the largest discrepancy between M1 and M2 could be $\sim$10 orders of magnitude for some species in the network. For models with an increase in grain temperature and decrease in diffusion barrier energy, as shown in Fig.~\ref{fig.overview_hist.20,0.3}, the averaged percentage within the double difference changed to 86\% and 64\% for the gas-phase species and the grain species, respectively. The largest difference in the logarithmic abundance ratio of M1 and M2 tends to decrease when the grain becomes warmer and the diffusion energy decreases.

\begin{figure*}[htbp]
\centering
\subfigure[Model: T$_d$ = 10 K, $E_{\rm dif}$/$E_{\rm des}$ = 0.5]{
    \includegraphics[scale=0.4]{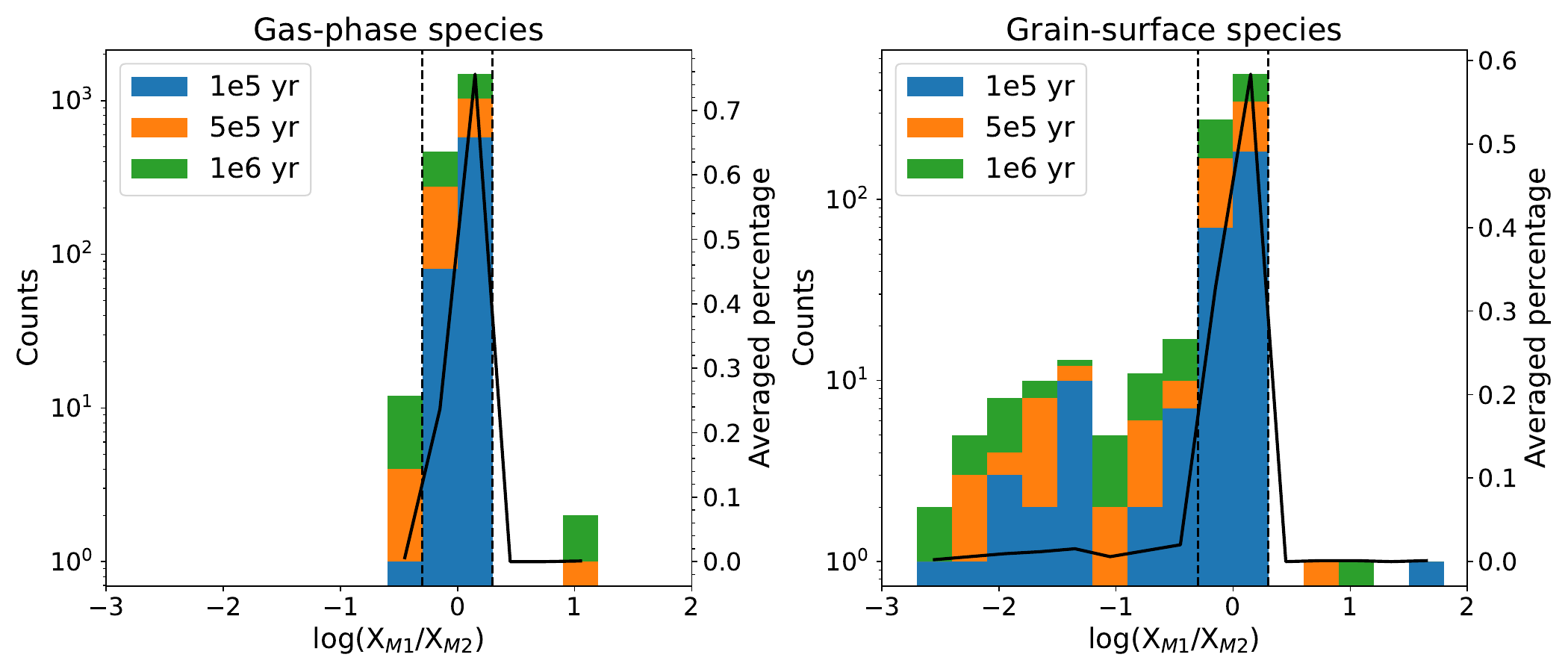}
}
\quad   
\subfigure[Model: T$_d$ = 20 K, $E_{\rm dif}$/$E_{\rm des}$ = 0.3]{
    \includegraphics[scale=0.4]{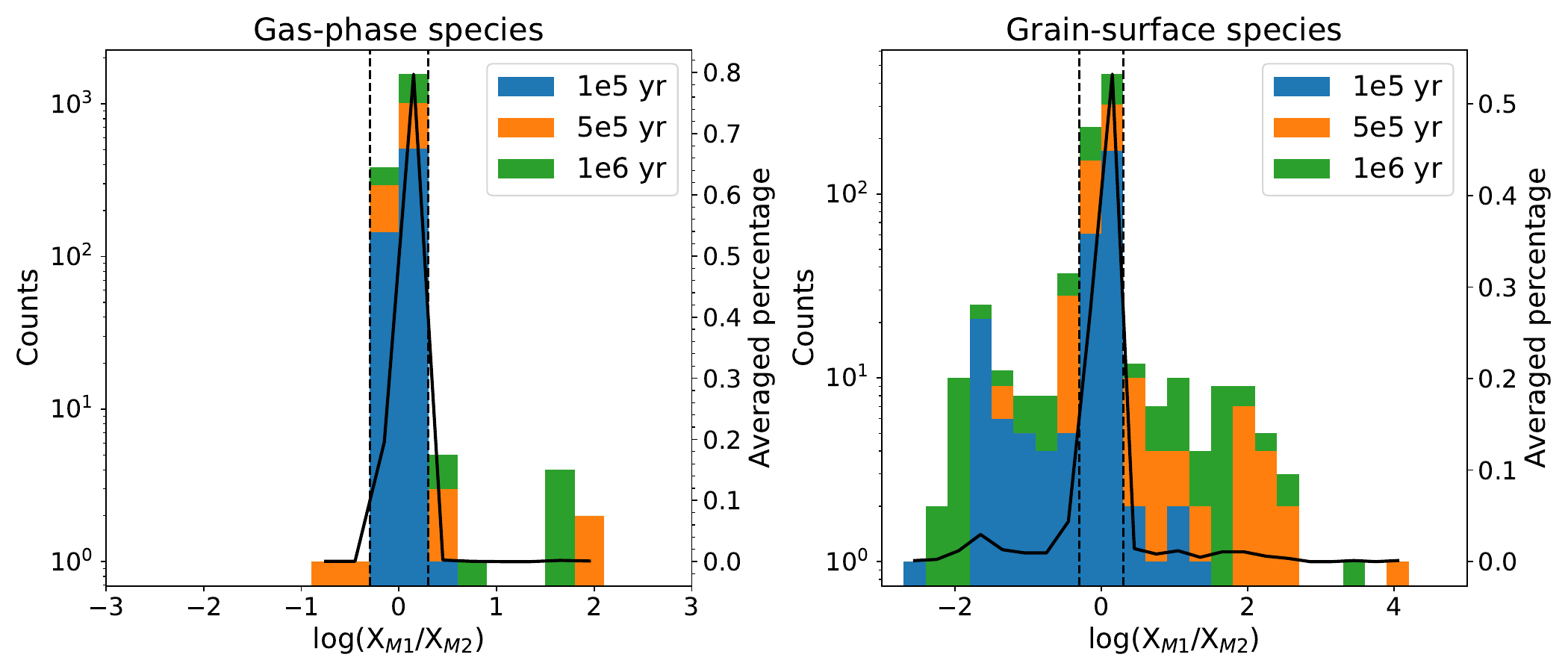}
}
\caption{Same as Fig.~\ref{fig.overview_hist}, but with the assumption that there is no difference when the abundance is lower than 10$^{-14}$ for a species in the two models.}
\label{fig.overview_hist_filter}
\end{figure*}

In Fig.~\ref{fig.overview_hist}, the high logarithmic abundance ratio may come from the species with a very low abundance. The observational abundance for a species is generally greater than 10$^{-12}$ under the cold molecular cloud conditions. When we assume that there is no difference when the abundance is lower than 10$^{-14}$ for a species in the two models, it might be expected that the difference between the two models should be smaller. Figure \ref{fig.overview_hist_filter} shows the statistical results for this assumption. 99\% of gas-phase species are within the double difference in both models, and the unaffected grain surface species also increased to over 80\%. Nevertheless, the abundance ratio for some grain surface species still exceeds one order of magnitude between M1 and M2. In the following, we focus on the analysis of the species with an abundance higher than 10$^{-14}$.

An increased grain temperature and decreased diffusion barrier energy increase the thermal diffusion rate of the grain species, which offsets the negative effect of the low value of the pre-factor used in the models on the chemistry. Generally, the reactants are accumulated due to their relatively slower diffusion rate on the grains in model M2 compared with M1, and the products are formed less correspondingly. As we show below, grain radicals are reserved in model M2, such as \ce{JOH}, \ce{JCH2OH}, and \ce{JCH3O} (here and hereafter, the prefix `J' stands for grain species. Its abundance is the sum of the abundances of its grain surface and ice mantle form). In contrast, the products of the reactions involving radicals are fewer, especially for grain COMs, such as \ce{JCH3CH2OH}, \ce{JCH3OCH3}, and \ce{JNH2CHO}.

\begin{figure*}[htbp]
\centering
\subfigure[Model: T$_d$ = 10 K, $E_{\rm dif}$/$E_{\rm des}$ = 0.5]{
    \includegraphics[scale=0.35]{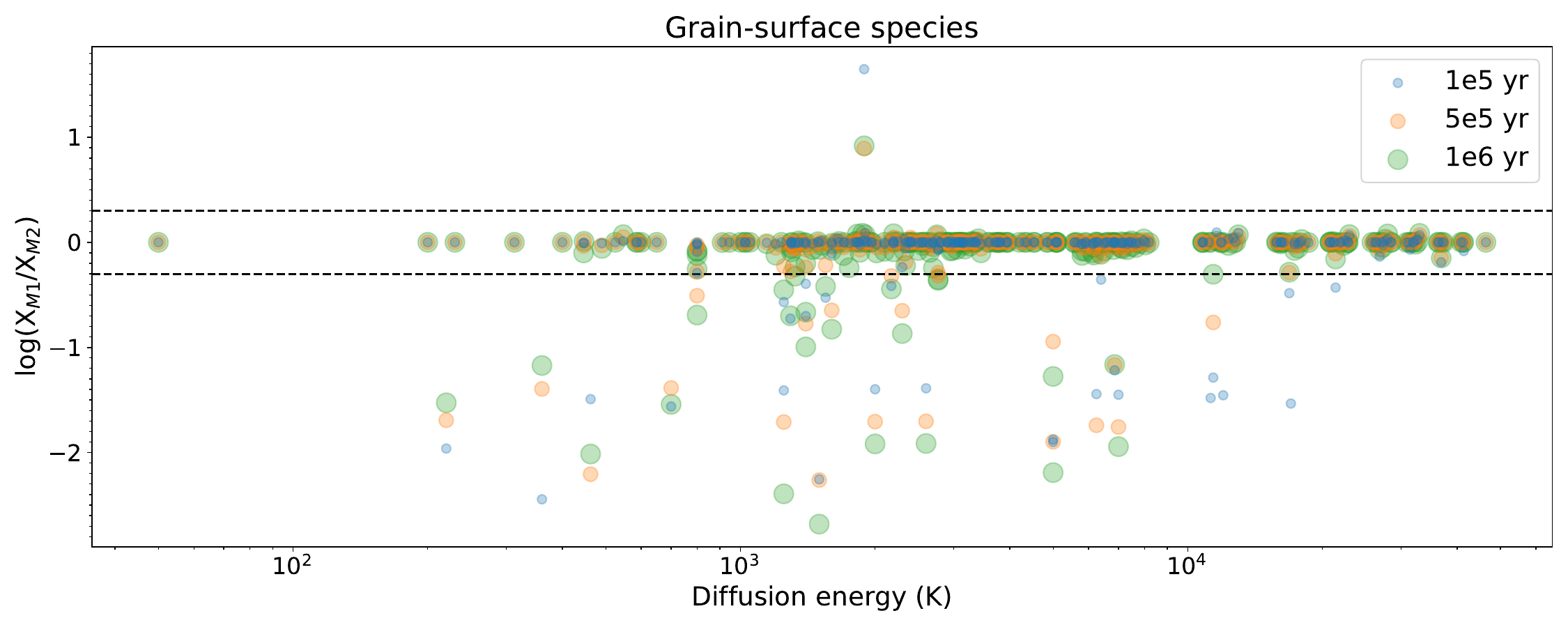}
    \label{fig.overview_Eb.10,0.5}
}
\quad  
\subfigure[Model: T$_d$ = 20 K, $E_{\rm dif}$/$E_{\rm des}$ = 0.3]{
    \includegraphics[scale=0.35]{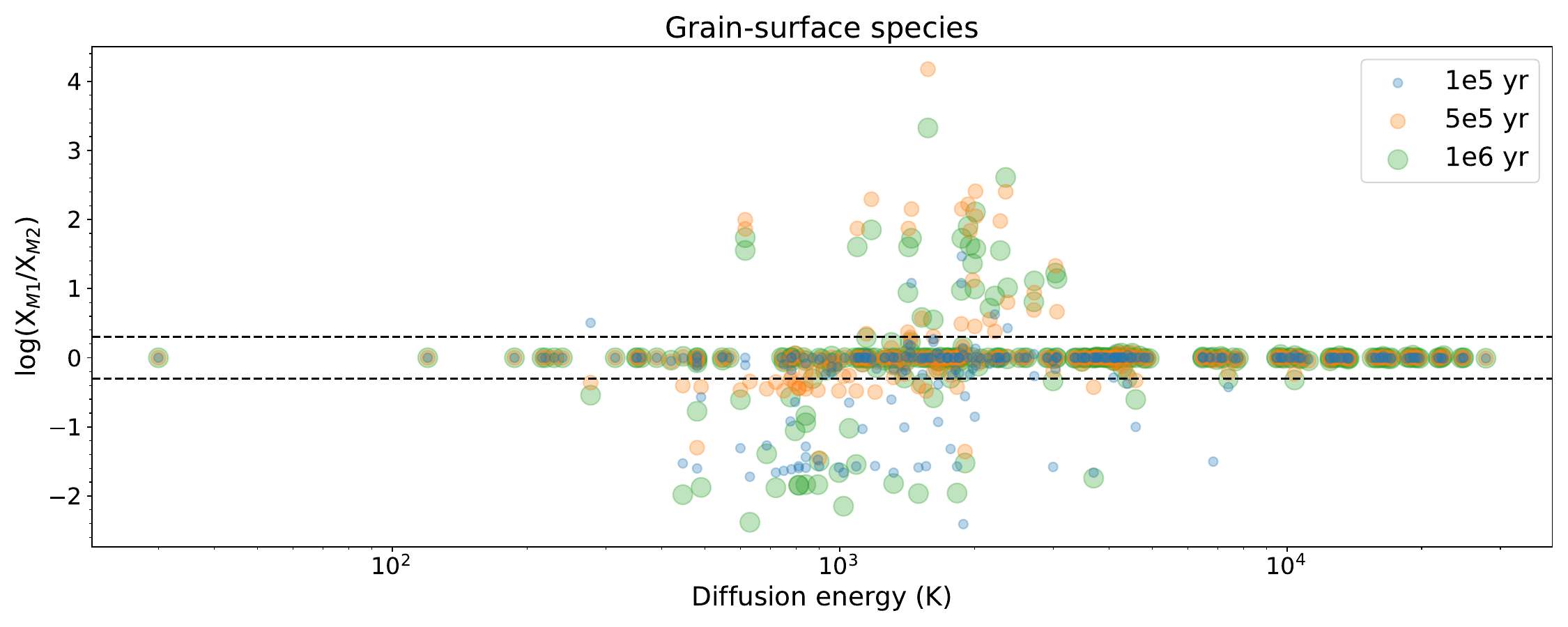}
    \label{fig.overview_Eb.20,0.3}
}
\caption{Distributions of the logarithmic molecular abundance ratio of M1 and M2 for the gas-phase species and grain species as a function of their diffusion energy at the three different evolutionary times. These results are under the assumption that there is no difference when the abundance is lower than 10$^{-14}$ for a species in the two models. Panel (a) shows model parameters with a grain temperature T$_d$ = 10 K and $E_{\rm dif}$/$E_{\rm des}$ = 0.5, and panel (b) shows T$_d$ = 20 K and $E_{\rm dif}$/$E_{\rm des}$ = 0.3. X$_{M1}$ and X$_{M2}$ represent the abundance of a species in models M1 and M2, respectively. The horizontal dashed lines plot the $\pm$0.3 difference of the logarithmic molecular abundance ratio of M1 and M2.}
\label{fig.overview_Eb}
\end{figure*}

Figure \ref{fig.overview_Eb} shows the distribution of the logarithmic abundance ratio of M1 and M2 as a function of the diffusion energy of the grain species. The diffusion energy is determined by the desorption energy times the ratio of $E_{\rm dif}$/$E_{\rm des}$, and a low diffusion energy leads to an increase in the diffuse rate of the species, so that we can expect that the most affected grain species are those with a relatively low diffusion energy. Some species are underproduced by M2 compared with M1, with an diffusion energy of about 3 $\times$ 10$^3$ K. These are mostly the grain COMs. For species that are overproduced by M2 with an diffusion energy higher than 10$^4$ K, they are mostly the species of long carbon-chains, JC$_n$H$_m$ with n > 4 and m = 0,1,2.

\subsection{Major ice components}
\begin{figure*}[htbp]
\centering
\includegraphics[scale=0.48]{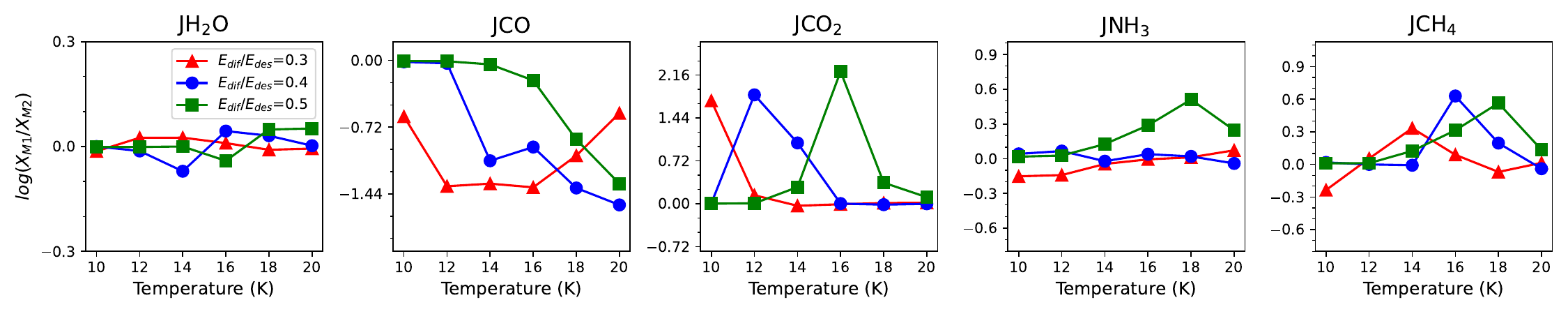}
\caption{Comparison of the logarithmic abundance ratio of M1 and M2 for the major ice components \ce{JH2O}, \ce{JCO}, \ce{JCO2}, \ce{JNH3}, and \ce{JCH4} for different grain temperatures and $E_{\rm dif}$/$E_{\rm des}$ ratios at 1 $\times$ 10$^5$ yr.}
\label{fig.major-ice}
\end{figure*}

Water is one of the major ice components. It can be produced by the successive hydrogenation of \ce{JO}. Model results show that the difference between M1 and M2 is smaller than 2 under varied grain temperatures and $E_{\rm dif}$/$E_{\rm des}$ ratios. The difference between M1 and M2 for \ce{JNH3} and \ce{JCH4} is smaller than 4, while these two species are formed by the hydrogenation with \ce{N} and \ce{C}, respectively. Figure \ref{fig.major-ice} shows the comparison of the logarithmic abundance ratio of M1 and M2 for \ce{JH2O}, \ce{JNH3}, and \ce{JCH4} for different grain temperatures and $E_{\rm dif}$/$E_{\rm des}$ ratios at 1 $\times$ 10$^5$ yr.

The formation of a species is determined by the concentrations of reactants and the rate constant of the reaction. The adsorption of gas-phase atomic \ce{O}, \ce{N}, \ce{C}, and \ce{H} onto the grain surface is the first step for the formation of \ce{JH2O}, \ce{JNH3}, and \ce{JCH4} ice. With a pre-factor value of 10$^{12}$, a hydrogen number density of 2 $\times$ 10$^4$ cm$^{-3}$, a grain temperature of 20 K, $E_{\rm dif}(H)$=230 K, and $E_{\rm dif}(O)$=800 K, we can estimate that the adsorption rate constant for \ce{H} and \ce{O} is about 10$^{-13}$ s$^{-1}$, and the diffusion rate constant for \ce{H} and \ce{O} on the grain surface is $\sim$10 and $\sim$4 $\times$ 10$^{-12}$ s$^{-1}$, respectively. The recombination rate constant for reaction \ce{JH + JOH -> JH2O} is proportional to the sum of the diffusion rate of reactants, which is mainly determined by the diffusion of \ce{H} in this reaction. The rate-determining step therefore is the adsorption of \ce{H} and \ce{O}. Because the adsorption processes are the same for models M1 and M2, the \ce{JH2O} abundance is similar in the two models.

\begin{figure*}[htbp]
\centering
\subfigure[\ce{JCO}]{
    \includegraphics[scale=0.35]{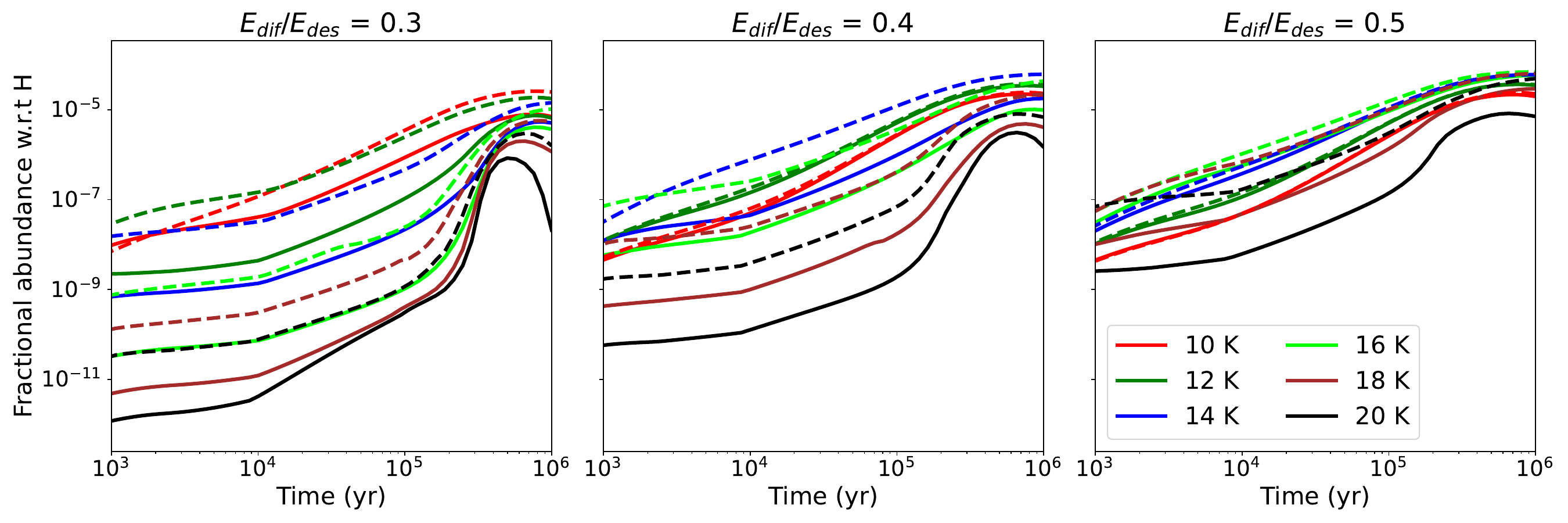}
    \label{fig.JCO}
}
\quad   
\subfigure[\ce{JCO2}]{
    \includegraphics[scale=0.35]{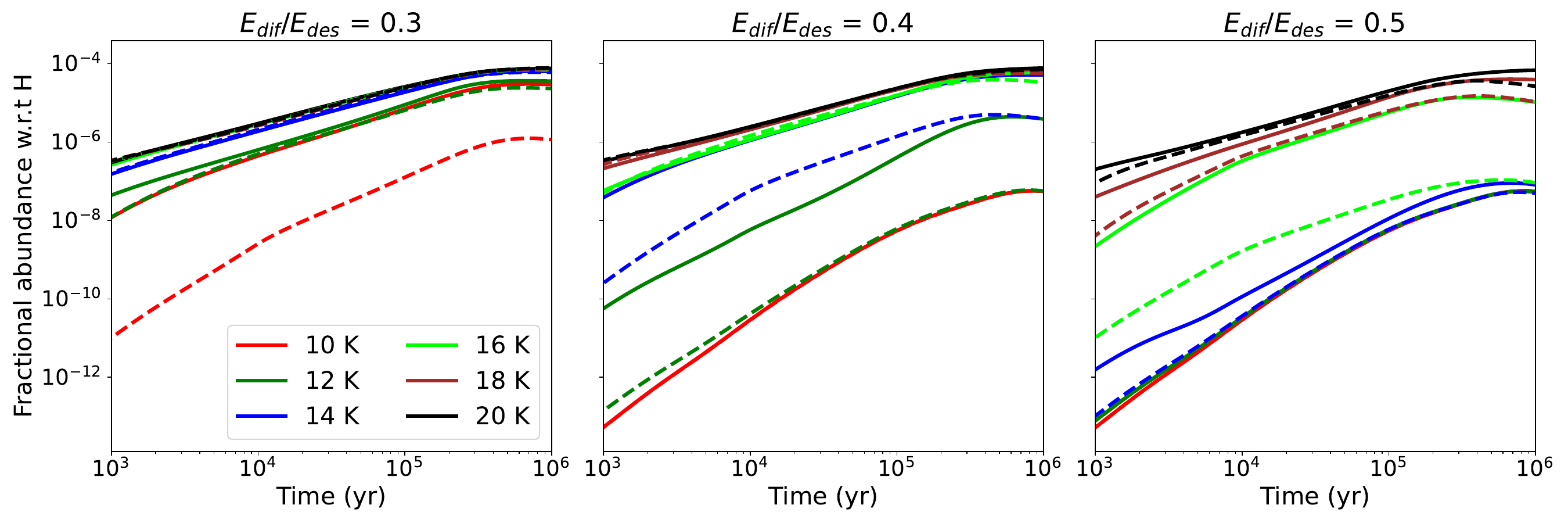}
    \label{fig.JCO2}
}
\caption{Fractional abundance of \ce{JCO} (upper panel) and \ce{JCO2} (lower panel) as a function of time for different grain temperatures and $E_{\rm dif}$/$E_{\rm des}$ ratios. The solid and dashed lines show models M1 and M2, respectively.}
\label{fig.JCO-JCO2}
\end{figure*}

For species that involve different types of reactants and multiple reaction pathways,  however, their evolution could be complicated. Figure \ref{fig.major-ice} also shows the comparisons for \ce{JCO} and \ce{JCO2}, and Fig.~\ref{fig.JCO-JCO2} shows the temporal evolution of their abundance. Figure \ref{fig.major-ice} shows that the abundance ratio of \ce{JCO} and \ce{JCO2} for M1 and M2 can vary by more than an order of magnitude depending on the grain temperature and the $E_{\rm dif}$/$E_{\rm des}$ ratio. With the increase in the ratio (i.e., less mobility of the species), a higher grain temperature is required to make the effect of the pre-factor more significant. \ce{JCO} participated in both the formation and destruction channels in the chemical reaction network, while \ce{JCO2} is less involved in destruction pathways, such as the reactions list from \ref{eq.JCO-reactant} to \ref{eq.JCO2-product}, so that \ce{JCO} could be accumulated because its thermal diffusion rate is lower in M2 than in M1. As a result, \ce{JCO} could be overproduced, while \ce{JCO2} could be underproduced in model M2.

\begin{chequation}
\begin{align}
\ce{JH + JCO  &-> JHCO}        \label{eq.JCO-reactant} \\
\ce{JH + JHCO &-> JCO + JH2}   \label{eq.JCO-product-1} \\
\ce{JO + JCO  &-> JCO2}        \label{eq.JCO-reactant-2} \\
\ce{JOH + JCO &-> JCO2 + JH}   \label{eq.JCO2-product}
\end{align}
\end{chequation}

\subsection{Minor ice components}
\begin{figure*}[htbp]
\centering
\includegraphics[scale=0.48]{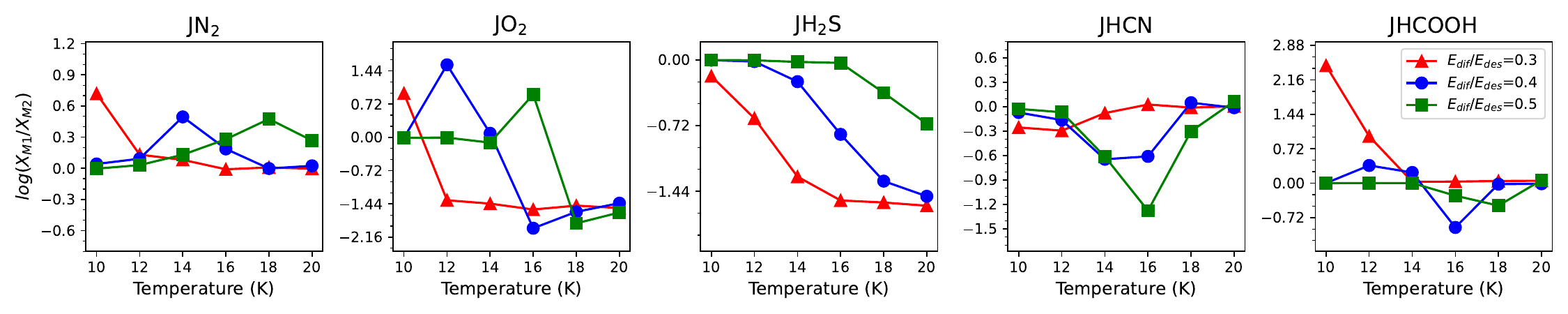}
\caption{Comparison of the logarithmic abundance ratio of M1 and M2 for five species of minor ice components, which are \ce{JN2}, \ce{JO2}, \ce{JH2S}, \ce{JHCN}, and \ce{JHCOOH} for different grain temperatures and $E_{\rm dif}$/$E_{\rm des}$ ratios at 1 $\times$ 10$^5$ yr.}
\label{fig.minor-ice}
\end{figure*}

Figure \ref{fig.minor-ice} shows five representative species of the minor ice components, which are \ce{JN2}, \ce{JO2}, \ce{JH2S}, \ce{JHCN}, and \ce{JHCOOH}. Their fractional abundance with respect to water ice is generally lower than a few percent. \ce{JN2} is one of the dominant N-bearing ice species. It is less affected by the changes in the pre-factor, while the maximum abundance difference between M1 and M2 is about five times the difference for specific grain temperatures and $E_{\rm dif}$/$E_{\rm des}$ ratios at 1 $\times$ 10$^5$ yr. The formation of \ce{JN2} is a chain of reactions following the adsorption of gas-phase atomic \ce{N} onto grains, the hydrogenation of \ce{JN}, the diffusion of \ce{JN}, and the recombination of \ce{JN} with \ce{JNH}. These processes are similar to the formation of \ce{JNH3} and \ce{JCH4}, except that the final recombination is \ce{JN} with \ce{JNH}. Therefore, the formation pathway for \ce{JN2} is similar to the formation pathway of \ce{JNH3} and \ce{JCH4}.

As for the chemistry of \ce{JH2S}, although \ce{JNH3}, \ce{JCH4}, and \ce{JH2S} are all involved in the hydrogenation processes, the difference for the abundance ratio of \ce{JH2S} between M1 and M2 is clear, where it is overproduced in M2 compared to M1. The main reason is that the initial sulfur abundance used in the model is 8 $\times$ 10$^{-8}$, which is the depleted abundance commonly used in cold cores. We found that when the initial sulfur abundance changes to 1.5 $\times$ 10$^{-5}$, which is the cosmic value for the sulfur element, the abundance ratio for \ce{JH2S} also shows a difference between M1 and M2 smaller than 2. The largest difference for the abundance ratio of \ce{JH2S} of M1 and M2 corresponds to a grain temperature of 20 K and $E_{\rm dif}$/$E_{\rm des}$ of 0.3, where the \ce{JH2S} abundance is 5 $\times$ 10$^{-13}$ in M1 and 2 $\times$ 10$^{-11}$ in M2.

\begin{figure*}[htbp]
\centering
\subfigure[]{
    \includegraphics[scale=0.35]{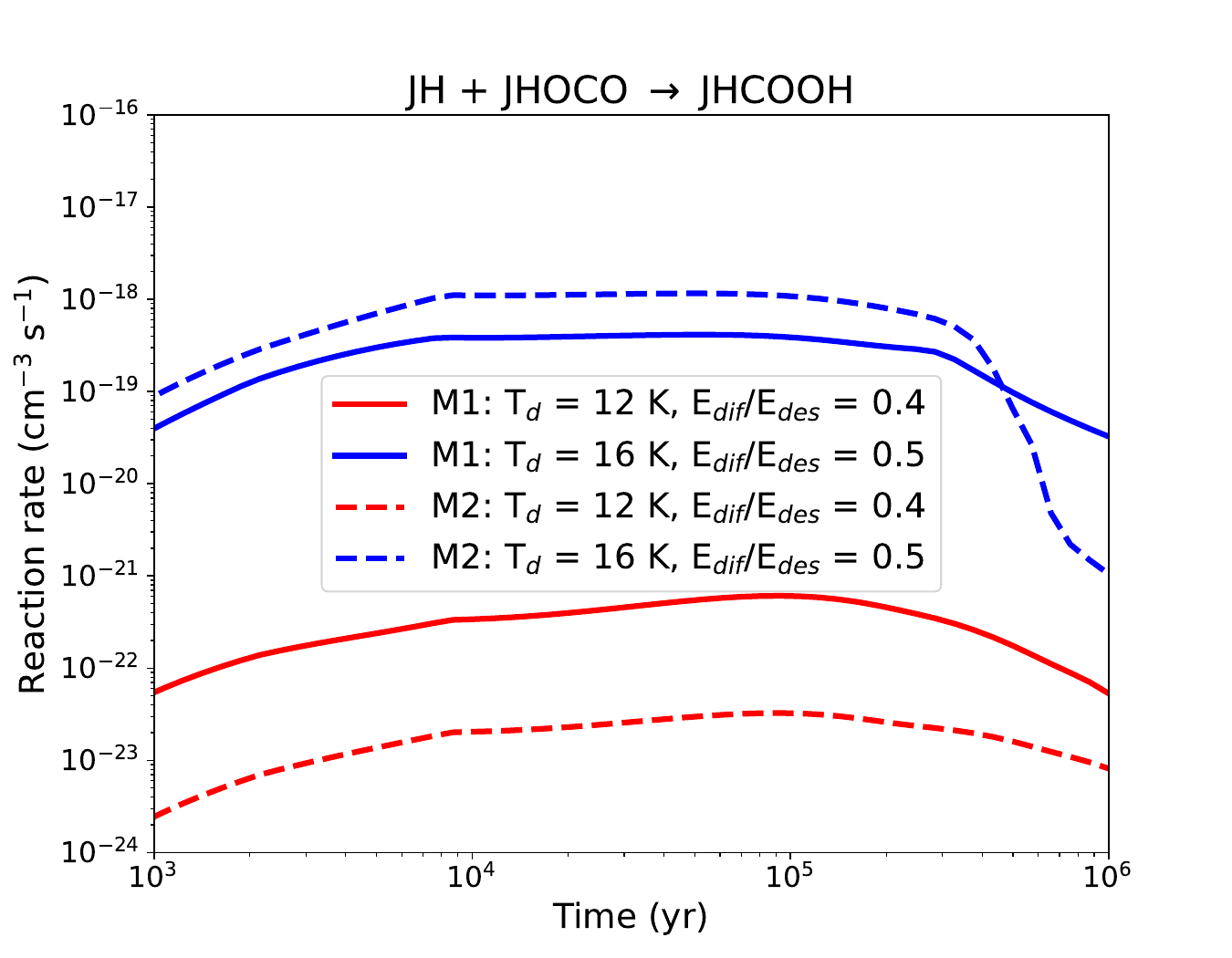}
    \label{fig.JHCOOH-rate}
}
\subfigure[]{
    \includegraphics[scale=0.35]{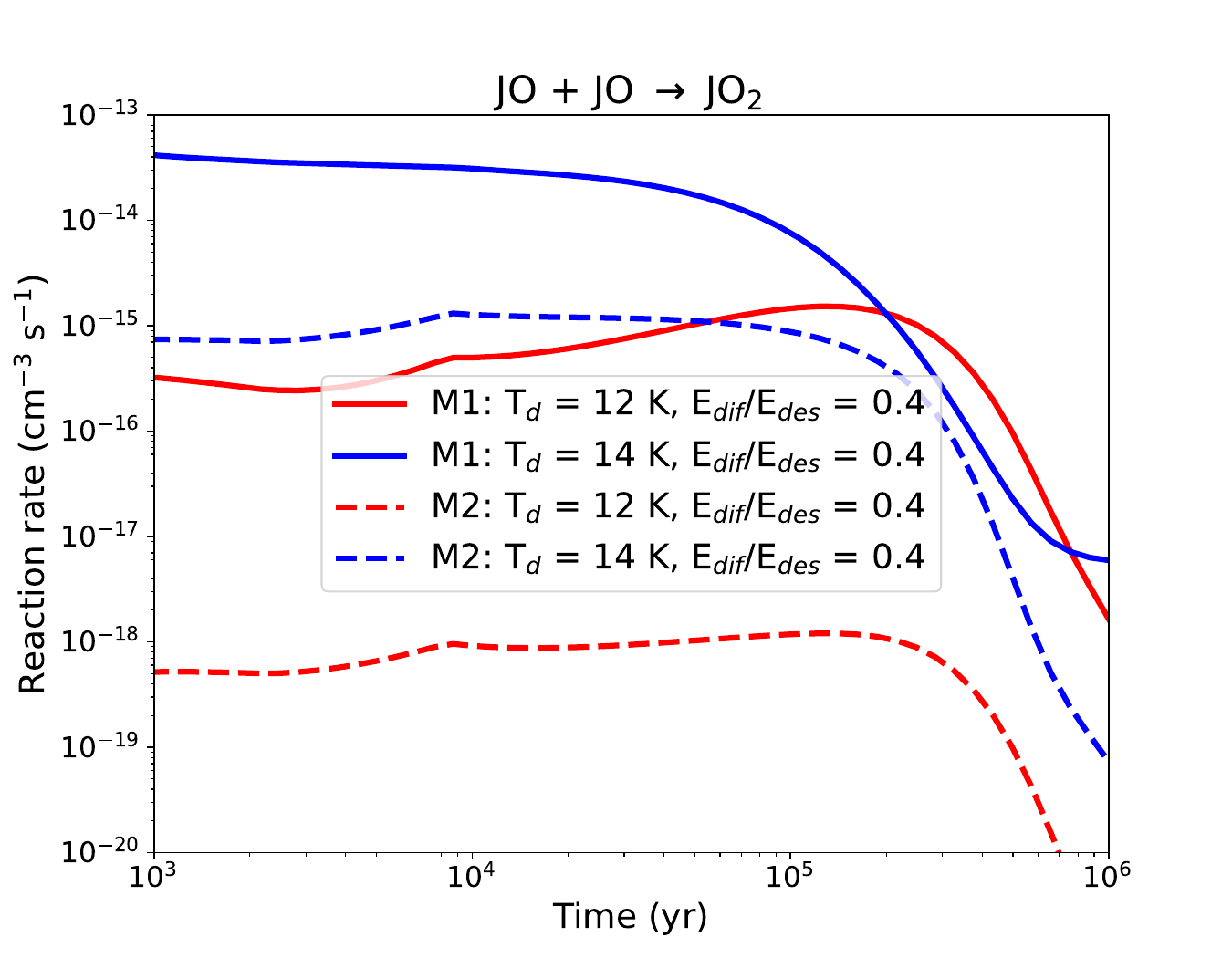}
    \label{fig.JO2-rate}
}
\caption{Reaction rate for \ce{JH + JHOCO -> JHCOOH} and \ce{JO + JO -> JO2} as a function of time. The solid and dashed lines show M1 and M2, respectively. The red and blue lines show the different parameters used in the models.}
\label{fig.minor-ice-rate}
\end{figure*}

\ce{JHCN} and \ce{JHCOOH} mainly participate in the formation reactions under cold dense molecular cloud conditions. It is therefore expected that they should be underproduced in M2 because their reactants are less mobile. However, modeling results show the opposite, as presented in Fig.~\ref{fig.minor-ice}. This is because their actual evaluation also depends on the number of reactants and on the reaction rate. The main formation pathway for \ce{JHCN} and \ce{JHCOOH} is \ce{JH2 + JCN -> JHCN + JH} and \ce{JH + JHOCO -> JHCOOH}, respectively. \ce{JHOCO} is formed by recombination of \ce{JOH} and \ce{JCO}. Thus, their formation depends on the mobility of the radical \ce{JCN} and \ce{JOH}. The following Fig.~\ref{fig.radicals} clearly shows that the logarithmic abundance ratios for these two radicals have their largest discrepancy between the two models under conditions of T$_d$ = 16 K and $E_{\rm dif}$/$E_{\rm des}$ = 0.5. Figure \ref{fig.JHCOOH-rate} presents the reaction rate for \ce{JH + JHOCO -> JHCOOH}, and it clearly shows an increase of some orders of magnitude of the reaction rate.

Finally, for the grain surface \ce{JO2}, unlike \ce{JCO} and \ce{JCO2}, \ce{JO2} could be both overproduced or underproduced in model M2 with respect to M1 by only a slight change in the grain temperature or the $E_{\rm dif}$/$E_{\rm des}$ ratio. This indicates its chemical sensitivity to these parameters. \ce{JO2} could serve as both a product and a reactant. The following reactions from \ref{eq.JO2-1} to \ref{eq.JO2-5} represent five of the most frequently contributed reaction pathways in the network. The total amount of \ce{JO2} depends on the competition of these reactions,

\begin{chequation}
\begin{align}
\ce{JO + JO   &-> JO2}           \label{eq.JO2-1} \\
\ce{JN + JO2H &-> JO2 + JNH}     \label{eq.JO2-2} \\
\ce{JH + JO3  &-> JO2 + JOH}     \label{eq.JO2-3} \\
\ce{JH + JO2  &-> JO2H}          \label{eq.JO2-4} \\
\ce{JO + JO2  &-> JO3}           \label{eq.JO2-5}.
\end{align}
\end{chequation}

\begin{figure*}[htbp]
\centering
\includegraphics[scale=0.35]{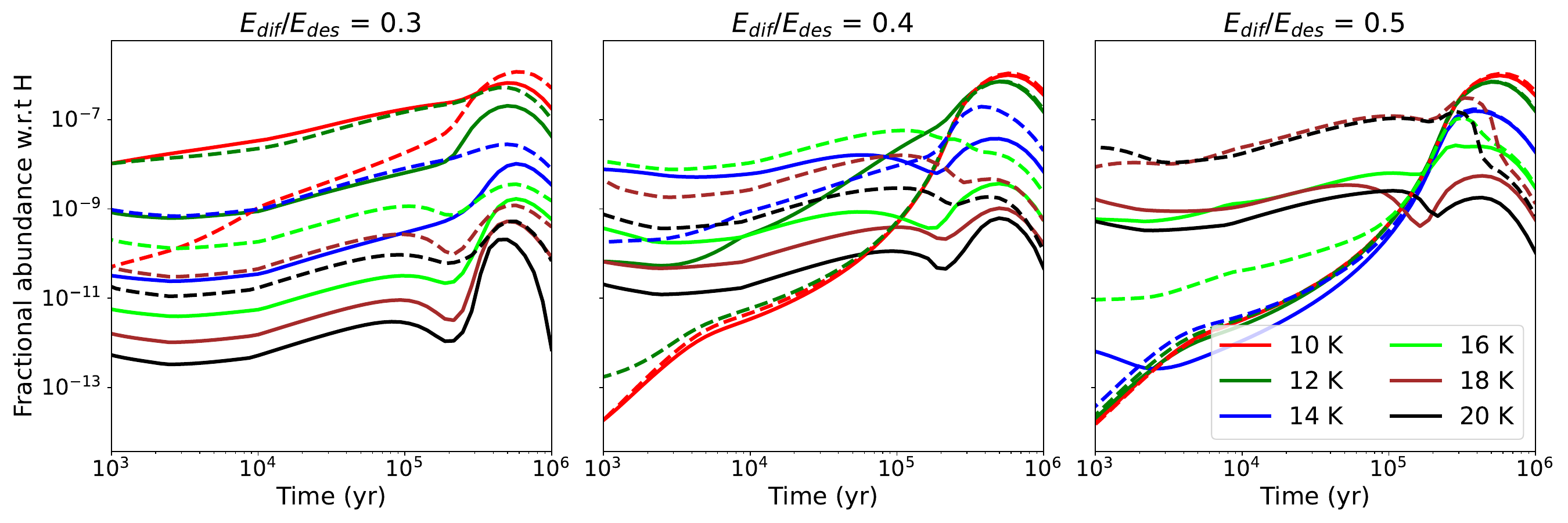}
\caption{Fractional abundance of \ce{JO2} as a function of time for different grain temperatures and $E_{\rm dif}$/$E_{\rm des}$ ratios. The solid and dashed lines show models M1 and M2, respectively.}
\label{fig.JO2}
\end{figure*}

Figure \ref{fig.JO2} shows the temporal evolution of \ce{JO2} at different grain temperatures and $E_{\rm dif}$/$E_{\rm des}$ ratios to reveal its complex temporal evolution as the parameter changes. Figure \ref{fig.JO2-rate} shows the reaction rate for the reaction \ref{eq.JO2-1} for M1 and M2 with a grain temperature of 12 and 14 K and an $E_{\rm dif}$/$E_{\rm des}$ ratio of 0.4. At a grain temperature of 12 K, the lower value of the pre-factor used in M2 could decrease the reaction rate, which could be expected in the early evolutionary times because of the proportional relation between the diffusion rate and the pre-factor. However, for a grain temperature of 14 K, the reaction rate in M2 quickly begins to approach the rate in M1. This indicates that the pre-factor is not the only parameter that contributes most to the difference in the chemistry between M1 and M2. Other parameters, such as the grain temperature, the diffusion barrier of the reactants, and the reaction pathways, could also play a critical role for the chemical evolution of the species. The temporal abundance depends on the competition of these factors.

\subsection{Grain radicals and COMs}
In addition to the major and minor grain species in the ice mantles, grain radicals and COMs are two important ingredients. Grain radicals participate in most of the grain surface reactions, and they could recombine into large species without barrier. Grain COMs are formed by the recombination of radicals, and they may be related with the origin of the gas-phase COMs. It is therefore required to determine the effect of the pre-factor on the radicals and COMs.

\begin{figure*}[htbp]
\centering
\includegraphics[scale=0.48]{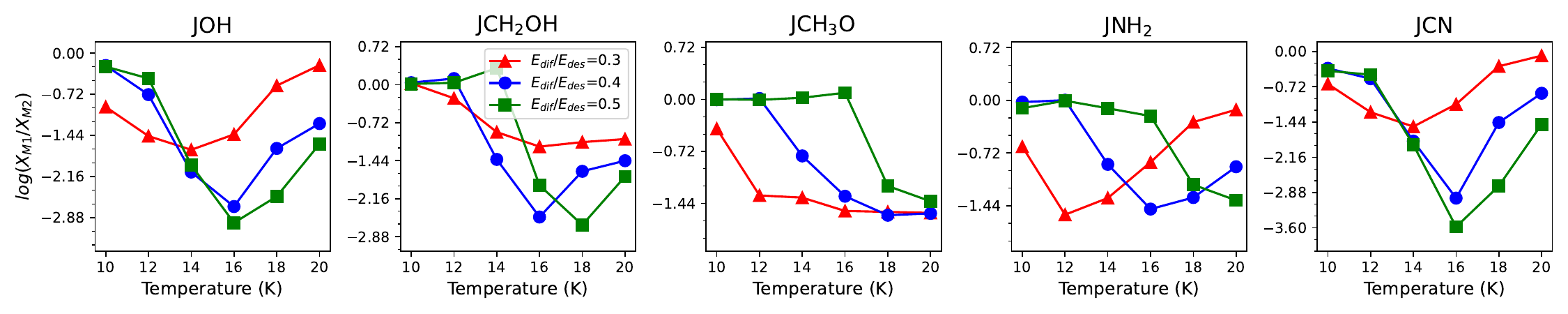}
\caption{Comparison of the logarithmic abundance ratio of M1 and M2 for five radicals in the ice mantles, which are \ce{JOH}, \ce{JCH2OH}, \ce{JCH3O}, \ce{JNH2}, and \ce{JCN}, for different grain temperatures and $E_{\rm dif}$/$E_{\rm des}$ ratios at 1 $\times$ 10$^5$ yr.}
\label{fig.radicals}
\end{figure*}

Figure \ref{fig.radicals} shows the comparison of the logarithmic abundance ratio of M1 and M2 for the grain radicals at 1 $\times$ 10$^5$ yr. For the radicals, the model result shows that all of them are overproduced or equally produced in model M2 compared to M1 at a given time. For \ce{JOH} and \ce{JCN}, it is overproduced in M2 at any given grain temperatures and $E_{\rm dif}$/$E_{\rm des}$ ratio. At T$_d$ = 16 K and $E_{\rm dif}$/$E_{\rm des}$ = 0.5, the abundance ratio difference is also maximized between the two models. This indicates that the effect of pre-factor reaches its maximum significance at the intermediate grain temperature, while an increase or decrease in the grain temperature could diminish this effect.

For other radicals, there is a clear trend that they are overproduced in M2 when the grain temperature increases and the $E_{\rm dif}$/$E_{\rm des}$ ratio decreases. For conditions of relatively low grain temperatures and high $E_{\rm dif}$/$E_{\rm des}$ ratios, they are equally produced in both models, except for \ce{JNH2} when $E_{\rm dif}$/$E_{\rm des}$ = 0.3. The reason for this is twofold. On the one hand, the lower the grain temperature and the higher the $E_{\rm dif}$/$E_{\rm des}$ ratio, the weaker the mobility of grain radicals, which causes the effect of the pre-factor to become negligible. On the other hand, compared with \ce{JOH} and \ce{JCN}, the formation steps for the hydrogenation of these radicals are longer.

It should be noted that the temporal evolution of the abundance changes as a function of time. The difference in the abundance ratio of M1 and M2 could therefore be different at longer evolutionary times. As mentioned previously, the diffusion rate of a surface species is negatively related to the diffusion barrier and positively related to the grain temperature. It therefore tends to be consumed at higher temperatures and lower diffusion barrier conditions. Thus, a lower value of the pre-factor in M2 could slow down the consumption of them, resulting in the accumulation of radicals in M2.

\begin{figure*}[htbp]
\centering
\includegraphics[scale=0.48]{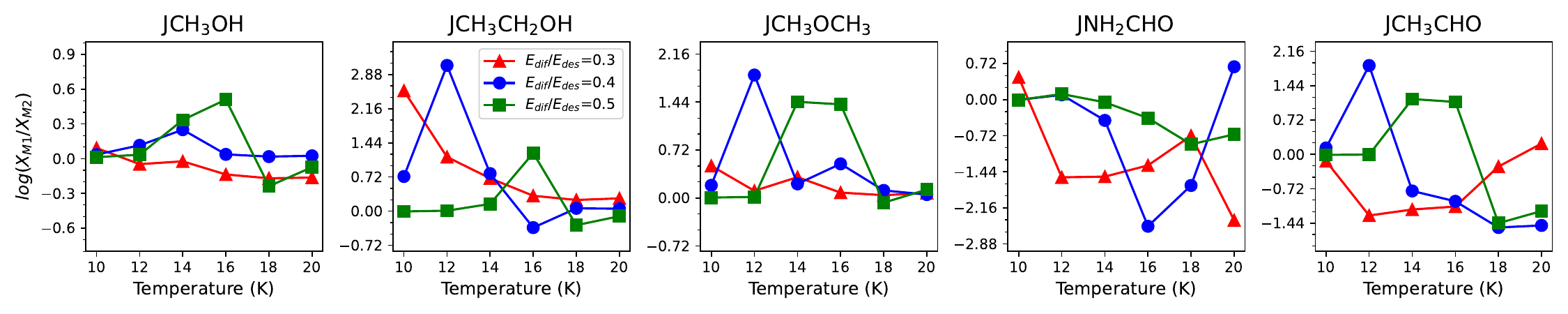}
\caption{Comparison of the logarithmic abundance ratio of M1 and M2 for five grain COMs, which are \ce{JCH3OH}, \ce{JCH3CH2OH}, \ce{JCH3OCH3}, \ce{JNH2CHO}, and \ce{JCH3CHO}, for  different grain temperatures and $E_{\rm dif}$/$E_{\rm des}$ ratios at 1 $\times$ 10$^5$ yr.}
\label{fig.JCOMs}
\end{figure*}

However, this is not the case for the grain COMs, as these species mainly participate in the formation pathways. The destruction of the grain COMs could be through the processes of desorption or dissociation (except for \ce{JCH3OH}, as discussed below), which are not dominant at typical cold dense molecular cloud conditions. Figure \ref{fig.JCOMs} shows the comparison of the logarithmic abundance ratio of M1 and M2 for the grain COMs at 1 $\times$ 10$^5$ yr. For \ce{JCH3OH}, the difference between M1 and M2 is smaller than 4. For \ce{JCH3CH2OH} and \ce{CH3OCH3}, which are produced by the radical-radical recombination, they are generally underproduced in M2 compared with M1 for most conditions.

\begin{figure*}[htbp]
\centering
\subfigure[\ce{JCH3OCH3}]{
    \includegraphics[scale=0.35]{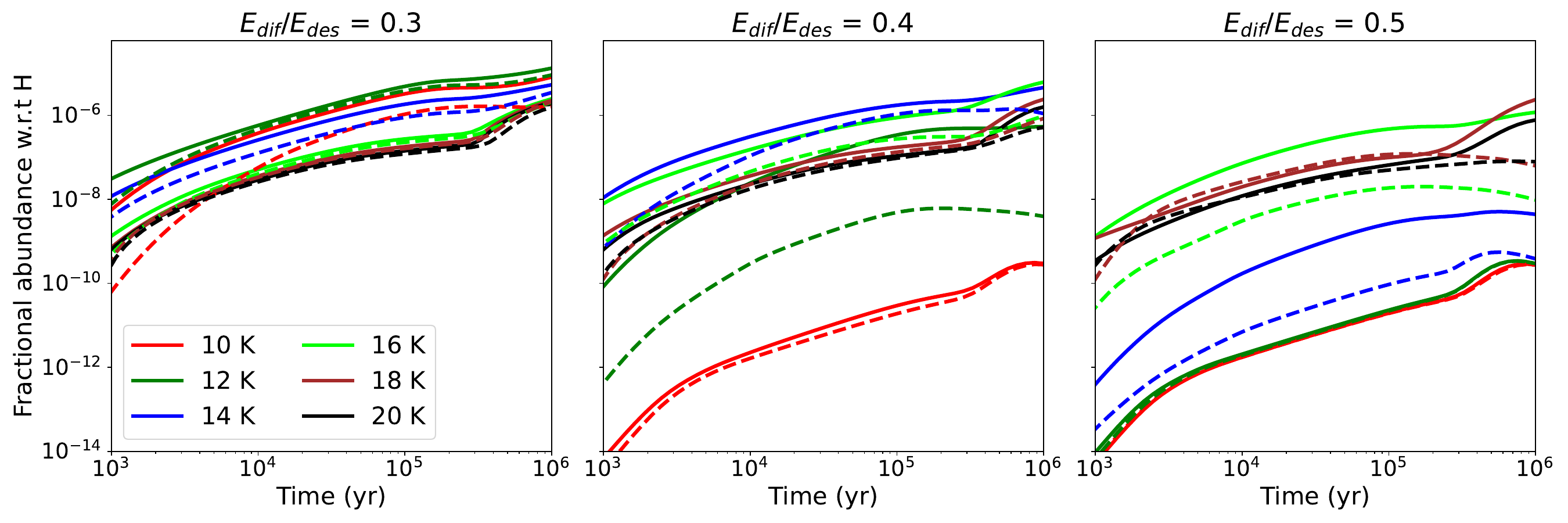}
    \label{fig.JCH3OCH3}
}
\quad   
\subfigure[\ce{JNH2CHO}]{
    \includegraphics[scale=0.35]{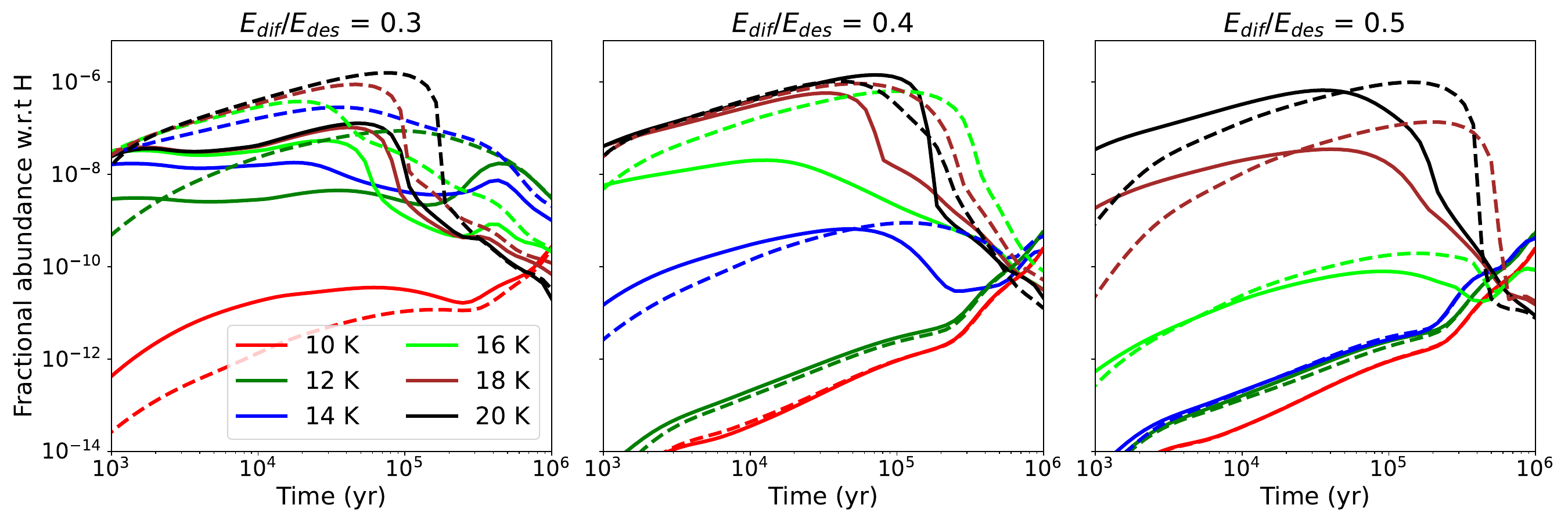}
    \label{fig.JNH2CHO}
}
\caption{Fractional abundances for the grain surface dimethyl ether (\ce{JCH3OCH3}) and formamide (\ce{JNH2CHO}) as a function of time for different grain temperatures and $E_{\rm dif}$/$E_{\rm des}$ ratios. The solid and dashed lines show models M1 and M2, respectively.}
\label{fig.JCOMs_abundance}
\end{figure*}

One of the most strongly affected species is the grain surface formamide, \ce{JNH2CHO}, where there are differences over 2 orders of magnitude between M1 and M2 at some given conditions. The main formation pathways for formamide in the network are the successive hydrogenation starting from \ce{JOCN}, where it is more abundant in M2, leading to the overproduced \ce{JNH2CHO} in M2. Figures \ref{fig.JCH3OCH3} and \ref{fig.JNH2CHO} show the abundance of grain surface dimethyl ether and formamide for models at different grain temperatures and $E_{\rm dif}$/$E_{\rm des}$ ratios, respectively. A lower diffusion barrier is favored for the formation of grain COMs. When the diffusion barrier becomes high enough, which decreases the mobility of the radicals, the increase in the grain temperature could further fill the deficiency of their formation caused by the high diffusion barrier. This is shown by the result for \ce{JCH3CHO}. \ce{JCH3CHO} can be formed by the hydrogenation of \ce{JCH2CHO} and \ce{JCH3CO}, and also by the radical-radical recombination between \ce{JCH3} and \ce{JHCO}. Thus, it can be efficiently produced in M2 under a low diffusion barrier and high grain temperature, but it is less efficiently produced under a high diffusion barrier and low grain temperature.

\begin{figure*}[htbp]
\centering
\subfigure[]{
    \includegraphics[scale=0.35]{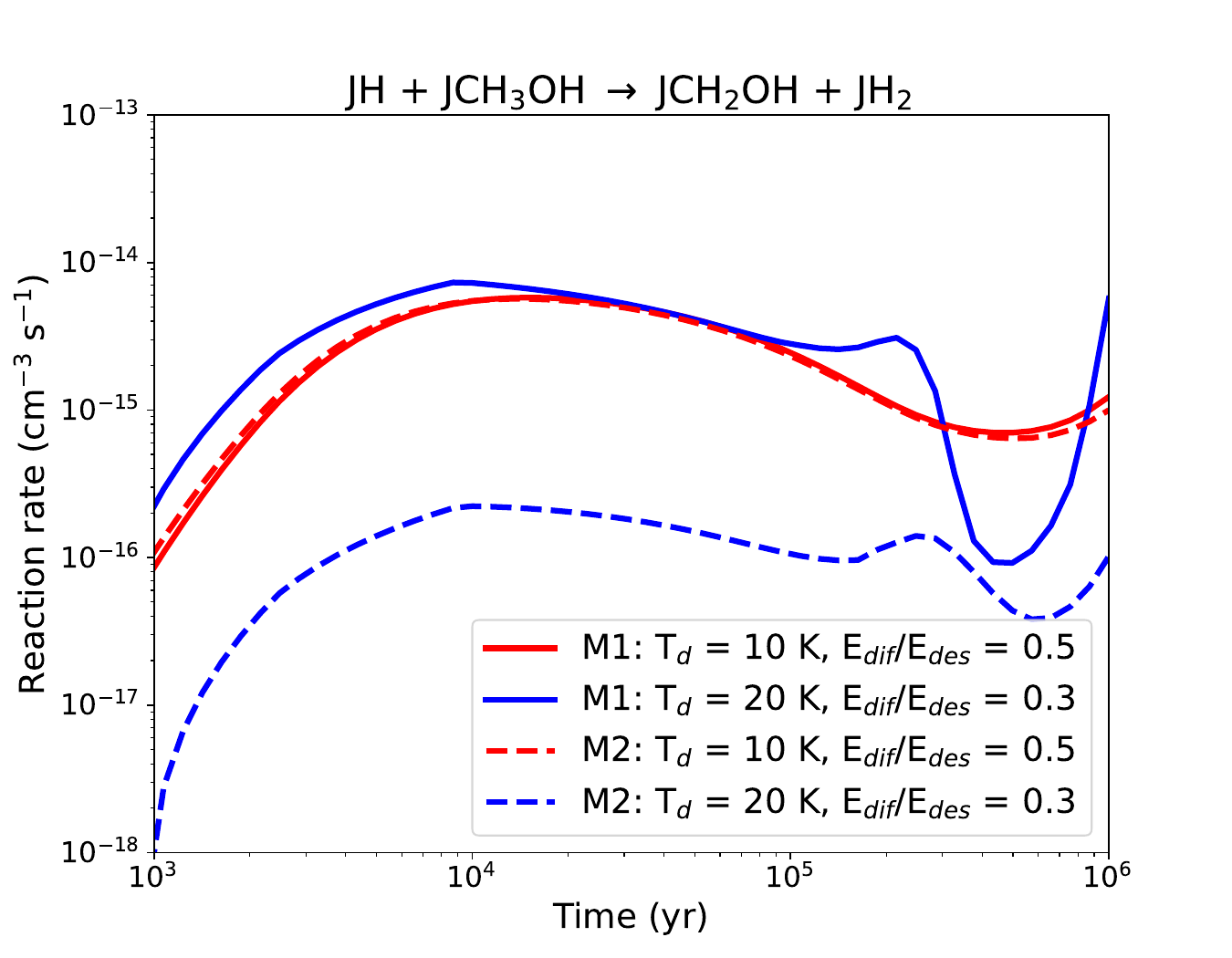}
    \label{fig.JCH3OH-rate}
}
\subfigure[]{
    \includegraphics[scale=0.35]{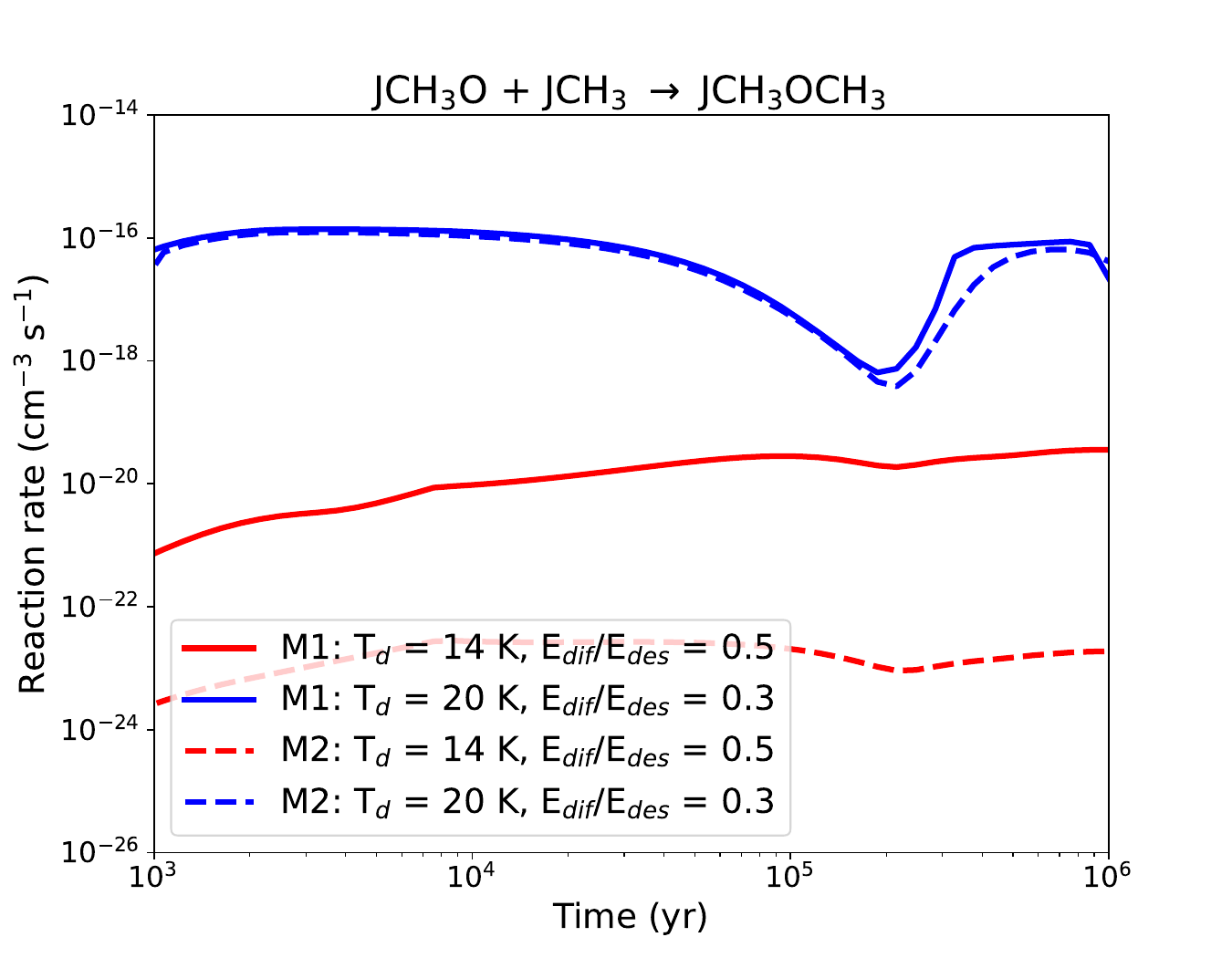}
    \label{fig.JCH3OCH3-rate}
}
\caption{Reaction rate for \ce{JH + JCH3OH -> JCH2OH + JH2} and \ce{JCH3O + JCH3 -> JCH3OCH3} as a function of time. The solid and dashed lines show M1 and M2, respectively. The red and blue lines show different parameters used in the models.}
\label{fig.JCOMs-rate}
\end{figure*}

Unlike other grain COMs, \ce{JCH3OH} could be hydrogenation dissociated through the following reaction \ref{eq.JCH3OH}. We plot the reaction rates of this reaction as well as the formation of \ce{JCH3OCH3} by radical recombination (reaction \ref{eq.JCH3OCH3}) in Fig.~\ref{fig.JCOMs-rate} for models with the given grain temperature and $E_{\rm dif}$/$E_{\rm des}$ ratio. At a grain temperature of 10 K and $E_{\rm dif}$/$E_{\rm des}$ of 0.5, the reaction rate of \ref{eq.JCH3OH} between M1 and M2 shows little difference. With the changes in grain temperature and $E_{\rm dif}$/$E_{\rm des}$, the difference between the two models is present. For the reaction \ref{eq.JCH3OCH3}, the largest difference between M1 and M2 comes from the model with the lower grain temperature. The reaction \ref{eq.JCH3OCH3} is barrierless, which means that its probability to occur is 1. The reaction \ref{eq.JCH3OH}, on the other hand, has an activation energy, so that its reaction probability also depends on the reaction-diffusion competition. Overall, the grain surface reactions are sensitive to the physical and chemical parameters,

\begin{chequation}
\begin{align}
\ce{JH + JCH3OH  &-> JCH2OH + JH2} \label{eq.JCH3OH} \\
\ce{JCH3O + JCH3 &-> JCH3OCH3}     \label{eq.JCH3OCH3}.
\end{align}
\end{chequation}

As a summary, for grain radicals, which primarily act as intermediate agents in the formation of COMs, the direct influence of the pre-factor is predominantly on the reaction diffusion rate. Therefore, using a lower pre-factor value in the models may result in an increased accumulation of these radicals within the ice mantles. The abundance of grain COMs, which primarily function as end products, is largely governed by the diffusion barrier of the reactants. When this diffusion barrier becomes sufficiently high, the temperature of the grain may emerge as the subsequent dominant factor, influencing the mobility of the reactants. Thus, a lower value of the pre-factor used in the models could scale down the rate of the reactions that form the grain COMs, leading to a weaker production of grain COMs.

\subsection{Simple gas-phase species and COMs}
Finally, in this section, we present the results of the effect of the pre-factor on the impact of the gas-phase species both for the simple molecules and COMs. The changes in the pre-factor initially only affect the grain surface species. However, grain surface species could desorb to the gas phase by various mechanisms, such as thermal desorption, chemical reactive desorption, UV photo-desorption, and cosmic-ray desorption. However, depending on the parameters used in the models, although there are $\sim$85\% of total gas-phase species that are within a double difference by the changes of the pre-factor, many gas-phase species are still largely affected by the pre-factor. For some species, the abundance difference between M1 and M2 could be over 3 orders of magnitude. These species are mainly COMs and their cations with molecular weights larger than 40 amu. However, it should be noted that the gas-phase abundances of these COMs are very low, even lower than 10$^{-30}$, which is far below the detection limit.

\begin{figure*}[htbp]
\centering
\includegraphics[scale=0.48]{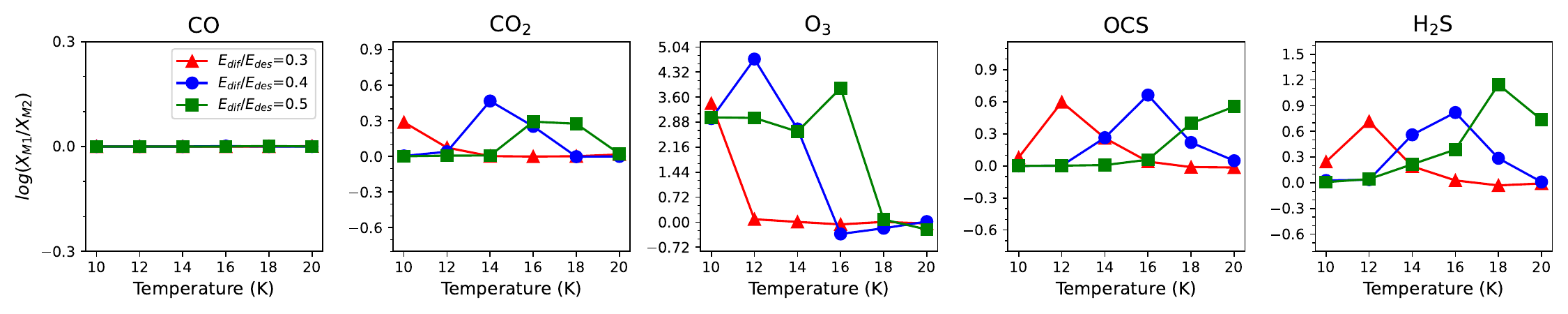}
\caption{Comparison of the logarithmic abundance ratio of M1 and M2 for five gas-phase simple species, which are \ce{CO}, \ce{CO2}, \ce{O3}, \ce{OCS}, and \ce{H2S}, for different grain temperatures and $E_{\rm dif}$/$E_{\rm des}$ ratios at 1 $\times$ 10$^5$ yr.}
\label{fig.gas-simple}
\end{figure*}

\begin{figure*}[htbp]
\centering
\includegraphics[scale=0.48]{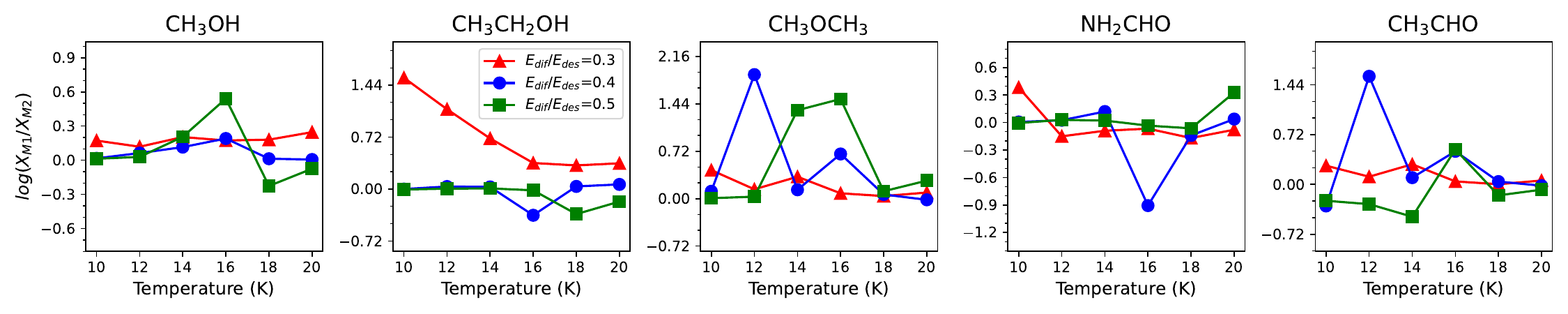}
\caption{Comparison of the logarithmic abundance ratio of M1 and M2 for five gas-phase COMs, which are \ce{CH3OH}, \ce{CH3CH2OH}, \ce{CH3OCH3}, \ce{NH2CHO}, and \ce{CH3CHO}, for different grain temperatures and $E_{\rm dif}$/$E_{\rm des}$ ratios at 1 $\times$ 10$^5$ yr.}
\label{fig.COMs}
\end{figure*}

Nevertheless, there are still notable differences for some of the gas-phase simple species and COMs. In the following sections, all the discussed gas-phase species are selected with an abundance higher than 10$^{-14}$, at least for most conditions. Figures \ref{fig.gas-simple} and \ref{fig.COMs} show the logarithmic abundance ratio of M1 and M2 for selected simple species and COMs, respectively. \ce{CO} is one of the most abundant gas-phase species, similar to the most abundant grain surface \ce{JH2O}. The changes in the pre-factor do not affect the abundance of \ce{CO} of M1 and M2 either. Other simple species, such as \ce{CO2}, \ce{OCS}, and \ce{H2S} in the figure, are underproduced by model M2, in which they are 3--6 times lower compared with M1 at 1 $\times$ 10$^5$ yr and given conditions. For \ce{O3}, the largest abundance difference could be over 3 orders of magnitude between M1 and M2. The primary factor contributing to the disparity in gas-phase abundance between M1 and M2 is chemically reactive desorption. This process means that a fraction of the molecules is desorbed into the gas phase as a result of the exothermic reaction that occurs during their formation on the grain surface.

Methanol is a basic and important COM. Figure \ref{fig.COMs} shows that the abundance difference for the gas-phase \ce{CH3OH} is smaller than two between M1 and M2 for different grain temperatures and $E_{\rm dif}$/$E_{\rm des}$ ratios, indicating that the changes in the pre-factor on the impact of \ce{CH3OH} are limited. For other COMs, they could be underproduced by M2 with one order of magnitude lower when compared with M1 for some conditions. For \ce{NH2CHO}, on the other hand, it could also be overproduced by M2 compared with M1 when the grain temperature is 16 K and \ce{E_{\rm dif}/E_{\rm des} is 0.4}. For cold molecular cloud conditions, the rates of the gas-phase synthesis of the COMs are inefficient, and they mostly originate from the grain surface followed by the various desorption mechanisms. The reactions for the formation of gas-phase methanol (\ref{eq.CH3OH-1}, \ref{eq.CH3OH-2}, \ref{eq.CH3OH-3}), dimethyl ether (\ref{eq.CH3OCH3}), and formamide (\ref{eq.NH2CHO}) are mainly contributed by the chemical reactive desorption. However, for the reactions to the formation of ethanol and acetaldehyde in the gas phase, not only the reactive desorption reactions (\ref{eq.CH3CH2OH-1}, \ref{eq.CH3CHO-1}), but the gas-phase electron dissociative recombination reactions (\ref{eq.CH3CH2OH-2}) and the neutral-neutral reaction induced by atomic \ce{O} (\ref{eq.CH3CHO-2}) contribute to their formation,

\begin{chequation}
\begin{align}
\ce{JH + JCH2OH     &-> CH3OH}        \label{eq.CH3OH-1} \\
\ce{JH + JCH3O      &-> CH3OH}        \label{eq.CH3OH-2} \\
\ce{JOH + JCH3      &-> CH3OH}        \label{eq.CH3OH-3} \\
\ce{JH + JCH3OCH2   &-> CH3OCH3}      \label{eq.CH3OCH3} \\
\ce{JH + JNH2CO     &-> NH2CHO}       \label{eq.NH2CHO} \\
\ce{JCH3 + JCH2OH   &-> CH3CH2OH}     \label{eq.CH3CH2OH-1} \\
\ce{C2H5OH2+ + e-   &-> CH3CH2OH + H} \label{eq.CH3CH2OH-2} \\
\ce{JH + JCH2CHO    &-> CH3CHO}       \label{eq.CH3CHO-1} \\
\ce{O + C2H5        &-> CH3CHO + H}   \label{eq.CH3CHO-2}.
\end{align}
\end{chequation}

\begin{figure*}[htbp]
\centering
\subfigure[\ce{CH3OCH3}]{
    \includegraphics[scale=0.35]{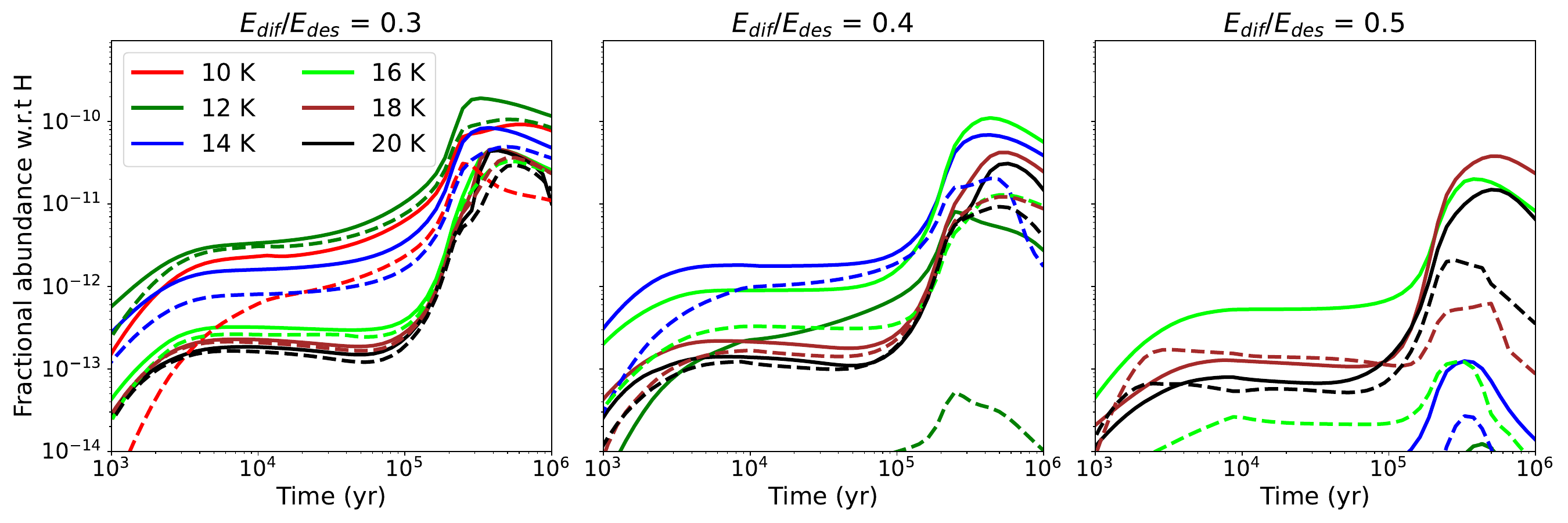}
    \label{fig.CH3OCH3}
}
\quad   
\subfigure[\ce{NH2CHO}]{
    \includegraphics[scale=0.35]{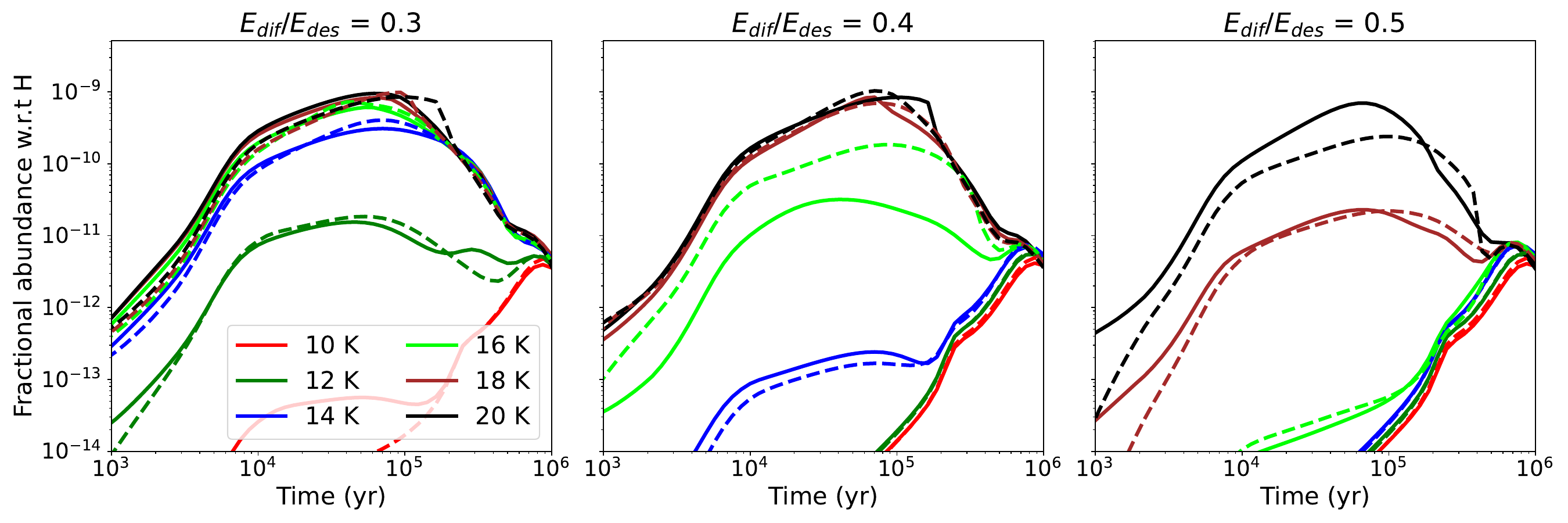}
    \label{fig.NH2CHO}
}
\caption{Fractional abundances for the gas-phase dimethyl ether (\ce{CH3OCH3}) and formamide (\ce{NH2CHO}) as a function of time for different grain temperatures and $E_{\rm dif}$/$E_{\rm des}$ ratios. The solid and dashed lines show models M1 and M2, respectively.}
\label{fig.COMs_abundance}
\end{figure*}

We finally show the abundance of dimethyl ether and formamide in Fig.~\ref{fig.COMs_abundance}. The peak abundance for dimethyl ether is about 10$^{-10}$, which is comparable with its observational value for cold core conditions, such as TMC-1. The peak abundance of formamide is about 10$^{-9}$ in warmer ($\sim$20 K) conditions. Most of the COMs are detected in hot cores or hot corinos with a high abundance. Thus, the effect of the pre-factor on the gas-phase COMs could be best revealed for hot core chemistry.

\section{Discussions} \label{sec.discussions}

\subsection{The effect of a different pre-factor on the chemistry}
There are two different types of pre-exponential factor. One factor is the desorption pre-exponential factor, and the other factor is the diffusion pre-exponential factor. The desorption pre-exponential factor can be used to determine the desorption energy and the thermal desorption rate of a species on the grain surface. The desorption energy determines at which grain temperature a species could desorb into the gas phase, which is important for the COMs chemistry in hot core or corinos (see the detailed discussions in \citet{Ligterink2023, Ferrero2022}). On the other hand, the diffusion pre-exponential factor is used to determine the diffuse energy and the thermal diffusion rate. Because the experiments are limited, the diffusion pre-exponential factor of only few molecules was constrained as presented in \citet{He2018, He2023}. The values of these diffusion pre-exponential factor is typically about 10$^9$ s$^{-1}$. However, in astrochemical models, the desorption and the diffusion pre-exponential factors are typically assumed to be the same according to the Hasegawa Eq.~(\ref{eq.Hasegawa}). The calculated values of the pre-exponential factor are generally about 10$^{12}$ s$^{-1}$, however, which is 3 orders of magnitude higher than the values determined by the experiments \citet{He2018, He2023}.

The rate of the thermal diffusion reaction in chemical models is directly influenced by the diffusion pre-factor. In this study, the value of the pre-exponential factor, as indicated by experiment data, was employed to assess its impact on the chemical processes. The results reveal that although over half of the total gas-phase and grain surface species, including abundant gas-phase \ce{CO} and grain surface water ice, remain unaffected by the pre-factor variations even after 10$^5$ yr of chemical evolution, the overall influence of a lower diffusion pre-factor on the chemistry is still significant and cannot be overlooked.

For the different types of species in the reaction network, that is,~the major and minor components of the ice mantles species, the radicals and COMs on the grain surface, and the simple species and COMs in the gas phase, the effect of the changes in the diffusion pre-factor on the abundance of part of them could vary strongly. For the grain surface species, the key impact species are those that serve as reactants, such as the radicals, or intermediate species that can diffuse and react with other species. These species generally have a molecular mass lower than 40--50 amu. For the large grain surface COMs, because they do not recombine with other species, their capacity to diffuse on the grain surface is negligible, however. 

Thus, the value of the pre-factor used in the chemical models should be carefully examined for a reliable prediction when compared with observational molecular abundance. More experiments for the simple and light species to constrain their diffusion pre-exponential factor could also be important to verify the range of their values.

\subsection{The significance of the diffusion pre-factor for different grain species}
We have shown that the diffusion pre-factor has a significant impact on the model predictions. We also identified the key species whose diffusion pre-factor would have a large impact on the model predicted abundances. We performed additional models for a selection of species. Compared with the previous model M2, where we assumed a uniform value of diffusion pre-factor of 10$^9$ s$^{-1}$, the value of the diffusion pre-factor changed to 10$^{12}$ s$^{-1}$ only for the selected species in the new model M3. To quantitatively represent the significance of the key species on the overall model results, we used the statistical value of the percentage that are within the double abundance difference between the two models at the time of 10$^5$, 5 $\times$ 10$^5$, and 10$^6$ yr as the criterion (see section \ref{sec.overview} for the details).

Table \ref{tab.key-species} shows the results for several grain surface species. The comparison is made under conditions of a grain temperature of 20 K and $E_{\rm dif}$/$E_{\rm des}$ ratio of 0.3, with the assumption that there is no difference when the abundance is lower than 10$^{-14}$ for a species in the two models. The changes in the diffusion pre-factor for \ce{JH}, \ce{JCO2} and \ce{JN2} have no impact on the model results because \ce{JH} is involved in the hydrogenation reactions, which are the quickest process compared with other reactions, while \ce{JCO2} and \ce{JN2} are not the agent of reactants so that their diffusion is not important. Grain radicals are important for the chemistry, but because of the short lifetime, it is difficult to measure their diffusion experimentally. Except for \ce{JCO}, whose diffusion has been measured experimentally by \citet{He2018}, we found that \ce{JCH3OH}, \ce{JH2CO}, and \ce{JNO} are among the key species that can significantly affect the model results. Further experiments to measure the diffusion properties of these ice species could be valuable.

\begin{table*}[htbp]
\centering
\caption{}
\label{tab.key-species}
\begin{tabular}{lll}
\hline
\hline
Species     & Gas-phase  & Grain surface \\
\hline
\ce{JH}     & 100\%      & 100\% \\
\ce{JN2}    & 100\%      & 100\% \\
\ce{JCH4}   & 100\%      & 99\%  \\
\ce{JHNCO}  & 100\%      & 99\%  \\
\ce{JO2}    & 99\%       & 98\%  \\
\ce{JCH3}   & 100\%      & 96\%  \\
\ce{JHCO}   & 99\%       & 94\%  \\
\ce{JCH3OH} & 100\%      & 93\%  \\
\ce{JH2CO}  & 98\%       & 89\%  \\
\ce{JCO}    & 99\%       & 86\%  \\
\ce{JNO}    & 98\%       & 83\%  \\
\hline
\end{tabular}\\
Notes\\
The effect of different grain surface species on the model is caused by changes in their diffusion pre-factor. The values are the percentages for the gas-phase and grain surface species that are within twice the abundance ratio difference of models M2 and M3. The lower the value, the greater the impact of the species on the model results.
\end{table*}

\section{Conclusions} \label{sec.summary}
We evaluated the effect of the diffusion pre-exponential factor (pre-factor) on the chemistry for cold molecular cloud conditions. The thermal diffusion rate of a grain surface species depends on the pre-factor, and the pre-factor shows a discrepancy between the experiments that perform on the ASW analogously to the interstellar ice conditions and the Hasegawa equation, which is commonly used in the chemical models. Models with varied grain temperatures and diffusion energy (E$_{\rm dif}$) to desorption energy ratios were also investigated to examine the effect of the diffusion pre-factor on the chemistry for these conditions. By distinguishing the diffusion pre-factor and the desorption pre-factor and using updated uniform diffusion pre-factor values of 10$^9$ s$^{-1}$ in the new models, we investigated the model results and compared the results from the reference model with the pre-factor determined by the Hasegawa equation. We grouped the species in the chemical reaction network into six types, which are the major and minor ice components species, the grain surface radicals and COMs, and the gas-phase simple species and COMs, to fully compare their logarithmic abundance ratio difference of the models. We summarize our conclusions below.

\begin{enumerate}
\item Statistically, over half of the total gas-phase and grain surface species are not affected by the changes in the diffusion pre-factor after 10$^5$ yr of chemical evolution. The percentage can increase to over 80\% when we assume that there is no difference when the abundance is lower than 10$^{-14}$ for a species in the model comparison. A lower grain temperature and higher diffusion barrier could minimize the impact of the pre-factor on the chemistry.

\item For the grain surface species that only act as the final products, such as \ce{JCO2}, and some COMs except for \ce{JCH3OH} or \ce{JNH2CHO}, they are underproduced in the new models compared with the reference models. The reason is lower mobility of the radicals that produce them. The species that participate in the reactants channels, such as the grain surface radicals, on the other hand, accumulate on the ice mantles because their mobility is weaker. Thus, they tend to be overproduced in the new models.

\item The effect of the changes in the diffusion pre-factor is more significant for the grain surface radicals because they participate in the recombination reactions. It is also significant at intermediate grain temperatures ($\sim$16 K) and high diffusion energy barriers conditions because a lower or higher grain temperature, or a lower E$_{\rm dif}$ could cancel out the impact of the changes in the pre-factor on the thermal diffusion rate.

\item Some gas-phase species could be affected by the desorption of grain surface species. However, the most abundant gas-phase \ce{CO} is not affected by the changes in the diffusion pre-factor, and the important COM in the gas-phase, \ce{CH3OH}, is affected only little. For those that are affected, most of them are generally underproduced in the new models.

\item The key species whose diffusion pre-factor significantly affects the model predictions was also evaluated. We recommend further experiments on the diffusion of ice species, such as \ce{CH3OH}, \ce{H2CO}, and \ce{NO}.

\end{enumerate}

\section{Acknowledgement}
We thank the referee for the helpful comments that greatly improved the manuscript. This work has been supported by the National Natural Science Foundation of China (NSFC) grant No.~11988101. L.-F.C.~ acknowledges the China Postdoctoral Science Foundation grant No.~2023M733271 and the Key Research Project of Zhejiang Lab (No.~2021PE0AC03). D.Q.~acknowledges the support by NSFC grant No.~12373026. J.H.~and T.H.~acknowledge the support from the European Research Council under the Horizon 2020 Framework Program via the ERC Advanced Grant Origins 83 24 28, and the support from the Vector Foundation. Di Li is a New Cornerstone Investigator. Y.W.~acknowledges the support by the NSFC grant No.~12303032, and the Natural Science Foundation of Jiangsu Province (Grant No.~BK20221163).


                
\end{document}